\documentclass[aps,nofootinbib,superscriptaddress,english,10pt]{revtex4}

\pdfoutput=1
\usepackage{etoolbox} 
\usepackage{lipsum} 

\makeatletter
\appto{\appendix}{%
  \@ifstar{\def\theequation@prefix{A.}}%
          {}%
}
\makeatother

\usepackage[latin1]{inputenc}
\usepackage{float}
\usepackage{latexsym}
\usepackage{graphicx}
\usepackage{amssymb}
\usepackage{cool}
\usepackage{dcolumn}
\usepackage{textcomp}
\usepackage{color}
\usepackage{amsmath}
\usepackage{amsfonts}
\usepackage{hyperref}
\usepackage[capitalize]{cleveref}

\def\beq{\begin{eqnarray}}
\def\eeq{\end{eqnarray}}

\def\({\left(}
\def\){\right)}

\def\lsim{\mathrel{\mathstrut\smash{\ooalign{\raise2.5pt\hbox{$<$}\cr\lower2.5pt\hbox{$\sim$}}}}}
\def\gsim{\mathrel{\mathstrut\smash{\ooalign{\raise2.5pt\hbox{$>$}\cr\lower2.5pt\hbox{$\sim$}}}}}

\begin{document}

\title{Unified Superfluid Dark Sector}
\author{Elisa G. M. Ferreira }\email{elisa.ferreira@mail.mcgill.ca}
\affiliation{Department of Physics, McGill University, Montr\'eal, QC, H3A 2T8, Canada}
\affiliation{Max-Planck-Institute for Astrophysics, Karl-Schwarzschild-Str. 1, 85741 Garching, Germany}
\author{Guilherme Franzmann}\email{guilherme.franzmann@mail.mcgill.ca}
\affiliation{Department of Physics, McGill University, Montr\'eal, QC, H3A 2T8, Canada}
\author{Justin Khoury}\email{jkhoury@sas.upenn.edu}
\affiliation{Center for Particle Cosmology, Department of Physics and Astronomy, University of Pennsylvania, Philadelphia, PA 19104, USA}
\author{Robert Brandenberger}\email{rhb@physics.mcgill.ca}
\affiliation{Department of Physics, McGill University, Montr\'eal, QC, H3A 2T8, Canada}

\date{\today}

\begin{abstract}
We present a novel theory of a unified dark sector, where late-time cosmic acceleration emerges from the dark matter superfluid framework.
The system is described by a superfluid mixture consisting of two distinguishable states with a small energy gap, such as the ground state and an excited state of dark matter. 
Given their contact in the superfluid, interaction between those states can happen, converting one state into the other. This long range interaction within the superfluid couples the two superfluid phonon species through a cosine potential motivated by Josephson/Rabi interactions. As a consequence of this potential, a new dynamics of late-time accelerated expansion emerges in this system, without the need of dark energy, coming from a universe containing only this two-state DM superfluid. Because the superfluid species are non-relativistic, their sound speeds remain suitably small throughout the evolution. We calculate the expansion history and growth of linear perturbations, and compare the results to $\Lambda$CDM cosmology. For the fiducial parameters studied here, the predicted expansion and growth function are close to those of $\Lambda$CDM, but the difference in the predicted growth rate is significant at late times. The present theory nicely complements the recent proposal of dark matter superfluidity to explain the empirical success of MOdified Newtonian Dynamics (MOND) on galactic scales, thus offering a unified framework for dark matter, dark energy, and MOND phenomenology. 
\end{abstract}

\maketitle

\section{Introduction}


The $\Lambda$ Cold Dark Matter ($\Lambda$CDM) model is currently the concordance model of cosmology. The model relies on two distinct dark components: dark energy (DE), in the form of a cosmological constant; and dark matter (DM), described as a cold pressureless fluid. Overall this model has been remarkably successful at matching a host of cosmological observations on large scales, {\it e.g.},~\cite{Ade:2015xua,Anderson:2013zyy,Abbott:2017wau}. The evidence that our universe is accelerating is abundant and comes from different types of observations~\cite{Anderson:2013zyy, Dodelson:2016wal}, with all pointing to the simplest explanation: the cosmological constant. Despite this striking agreement, however, a number of nagging tensions have emerged in the data on linear scales. These include the lack of power at large angles in the cosmic microwave background (CMB) angular correlation function~\cite{Schwarz:2015cma}, and the largely debated tension between estimates of the Hubble parameter from the CMB and local measurements~\cite{Riess:2018byc, Macaulay:2018fxi}. 

Together with these observational tensions, the $\Lambda$CDM model also presents serious theoretical challenges. The smallness of the cosmological constant is vexing because of its radiative instability under quantum corrections, {\it e.g.},~\cite{Padilla:2015aaa}. This is the cosmological constant problem. Furthermore, there is no compelling explanation for the coincidence that the densities of DM and DE are of the same order at the present time, despite having very different evolution. This is the coincidence problem. These problems have motivated the search for alternative explanations for the nature of DE. Many alternative models can be found in the literature, ranging from dynamical DE models (reviewed in~\cite{Copeland:2006wr}) and modifications to General Relativity (GR) (reviewed in~\cite{Joyce:2014kja}). However none of the models on the market address the cosmological constant problem. On the contrary they often exacerbate things by introducing new fine tunings.

Meanwhile, the idea that DM is described by a cold, collisionless fluid has proven successful at explaining large-scale observables. 
On small, galactic scales, the scorecard of CDM is murkier and has been the subject of active debate, {\it e.g.},~\cite{DelPopolo:2016emo,Bullock:2017xww}. 
Some of the small-scale challenges include the internal structure of dwarf galaxies in the Local Group~\cite{BoylanKolchin:2011de,BoylanKolchin:2011dk} and possibly beyond~\cite{Papastergis:2014aba}, the vast planar structures seen around the Milky Way and Andromeda~\cite{Pawlowski:2012vz,Pawlowski:2013kpa,Pawlowski:2013cae,Ibata:2013rh}, and other issues related to dynamical friction between baryons and a live dark matter (DM) halo, {\it e.g.},~\cite{Debattista:1997bi,Berrier:2016gki}. 

Arguably, the most pressing challenge is the mass discrepancy acceleration relation (MDAR). The MDAR is a remarkably tight correlation between the dynamical gravitational acceleration inferred from rotation curves and the gravitational acceleration due to baryons only, as inferred from the distribution of stars and gas~\cite{McGaugh:2016leg,Lelli:2017vgz}. In plain DM parlance, the MDAR implies that by looking only at the baryon mass distribution, one can infer the DM density profile at every radius in the galaxy, even in galaxies where baryons are everywhere subdominant. At large distances in the disk, the MDAR implies the baryonic Tully-Fisher relation (BTFR)~\cite{McGaugh:2000sr,McGaugh:2011ac}, which relates the total baryonic mass to the asymptotic/flat rotation velocity with remarkably small scatter~\cite{LelliBTFR}.  In the central region, the MDAR implies the correlation between stellar and dynamical surface densities in disk galaxies~\cite{Lelli:2016uea}.

The MDAR was of course predicted by Milgrom over thirty years ago~\cite{Milgrom:1983ca}. Milgrom's law states that the total gravitational acceleration $a$ is approximately $a_{\rm N}$ in the regime $a_{\rm N}\gg a_0$, and approaches the geometric mean $\sqrt{a_{\rm N} a_0}$ whenever $a_{\rm N}\ll a_0$, where $a_{\rm N}$ is the usual Newtonian gravitational field generated from the observed distribution of baryonic matter alone, and the transition acceleration scale is $a_0 \sim 10^{-8}{\rm cm}/{\rm s}^{2}$. Historically this empirical law was proposed as an alternative to DM known as MOdified Newtonian Dynamics (MOND). By now this possibility seems unlikely, given aforementioned large body of evidence for DM as a collisionless fluid on cosmological and cluster scales.

The model of DM superfluidity~\cite{Berezhiani:2015bqa,Berezhiani:2015pia,Khoury:2016ehj,Berezhiani:2017tth} proposes to successfully marry the phenomenological success of the $\Lambda$CDM model on cosmological scales with that of MOND on galactic scales. (See also~\cite{Khoury:2016egg,Hodson:2016rck,Addazi:2018ivg,Cai:2017buj,Fan:2016rda,Alexander:2018fjp,Hossenfelder:2018iym}.) Through the well-known physics of superfluidity, the collisionless fluid and ``modified gravity" phenomena are two sides of the same coin, representing different phases of the underlying DM theory. The DM candidate in this case consists of axion-like particles with sufficiently strong self-interactions such that they thermalize in galaxies. With $m\sim {\rm eV}$, their de Broglie wavelengths overlap in the (cold and dense enough) central region of galaxies, resulting in Bose-Einstein condensation into a superfluid phase. 

The superfluid nature of DM dramatically changes its macroscopic behavior in galaxies. Instead of behaving as individual collisionless particles,
the relevant low energy excitations are phonons. It is crucial that the DM phonons must couple to baryons in a way that explicitly
(but softly) breaks the global $U(1)$ symmetry of the superfluid. As a result, the DM phonons mediate a long-range force that effectively modifies gravity. 
(See~\cite{Famaey:2017xou} for a recent proposal for obtaining the MDAR from particle interactions between DM and baryons.)
For a particular choice of the superfluid equation of state, the resulting phonon effective Lagrangian is similar to the MOND scalar field theory~\cite{Bekenstein:1984tv}.
Remarkably, this phonon effective theory is strikingly similar to that of the Unitary Fermi Gas ({\it e.g.},~\cite{Giorgini:2008zz}), which has generated
much excitement in the cold atom community in recent years. In recent work~\cite{Sharma:2018ydn}, the finite-temperature equation of state of DM superfluids was computed using a self-consistent mean-field approximation.

\subsection{A unified description of superfluid DM and DE}

A natural question is whether late-time cosmic acceleration can also emerge in the DM superfluid framework, as yet another manifestation of the same underlying substance as DM and MOND. In this paper we will argue that this is indeed possible. We propose a model where the DM consists of a mixture of two superfluid species, arising from two distinguishable states of DM separated by a small energy gap. Although we will not commit to a particular microphysical realization, a natural possibility is that DM is made of dark atoms~\cite{Goldberg:1986nk,Hodges:1993yb} with hyperfine splitting of the ground state. In this case, the two-state superfluid framework can be thought of as a refinement of the previous model that now takes into account the atomic structure of DM.

Since DM particles in these different states are in contact with each other in the superfluid, it is natural to assume that they can scatter via contact interactions. Such interactions lead to a conversion between the different states (or species) of DM particles of the superfluid. This interaction is similar to a Rabi coupling or an internal Josephson interaction {\it e.g.},~\cite{Josephson,Mahan}. The presence of the two states and their interaction has consequences for the collective behavior of the superfluid, altering some of its properties, such as changing the ground state of this system. The study of such systems and its properties is a very active field in condensed matter physics, being studied theoretically~\cite{Tommasini} (see~\cite{Usui,Cappellaro,Bornheimer} as examples of recent developments) and experimentally (\textit{e.g.},~\cite{lab1,Zibold}). We are interested in understanding the implications for cosmology of DM being described by a two-component interacting superfluid.
 
 A superfluid is described in terms of its collective variable, namely the condensate wavefunction describing the entire superfluid. At low energy, the relevant degree of freedom is the phase of this wavefunction, which is the Goldstone mode for the spontaneous $U(1)$ breaking. Its excitations are phonons (sound waves). In a two-state superfluid mixture, each superfluid specie has different phonon excitations, $\theta_1$ and $\theta_2$, describing two different sound waves propagating through the system. The contact interaction between particles in the two different states results in an oscillatory potential that depends on the difference of the phonon phases: 
 \beq
V(\theta_2-\theta_1 + \Delta E \,t) = M^4\cos^2 \left(\frac{\theta_2-\theta_1+ \Delta E \,t}{2}\right)\,,
\label{cosinepotential_intro}
\eeq
where $\Delta E$ is the small energy gap between the two distinguishable states of DM. (We work in natural units throughout the paper, unless stated otherwise.) This potential explicitly breaks the individual global $U(1)$ symmetries down to a diagonal $U(1)$ subgroup. Physically, this encodes the fact that particles of one species can be converted into the other, but the total number of particles is conserved. Such interaction potentials are ubiquitous in systems with multiple superfluid/superconducting species. Through this potential energy, a multi-state interacting DM component can give rise to new emergent dynamics coming from the collective behavior of this superfluid --- a dynamical phase of cosmic acceleration. This phase will occur at the present time provided that $M^4$ is of order $\sim {\rm meV}^4$. As we will see, the energy gap $\Delta E$ must be smaller than $H_0$, so that the accelerating phase is approximately de Sitter. 

Despite the similarity with axions and pseudo-Nambu Goldstone Boson (pNGB) models~\cite{Frieman:1995pm,Kaloper:2005aj}, there are important differences. Firstly, the kinetic terms for the angular fields $\theta_i$ are non-canonical. They are described by non-linear $P(X)$ functions, which encode the superfluid equation of state. Because of this difference in kinetic term, we will see that the DM phase is obtained without oscillations of the field around the bottom of the potential, thus avoiding potential instabilities~\cite{Johnson:2008se}. Despite this difference in kinetic term, our model also requires super-Planckian decay constants to drive cosmic acceleration, like in pNGB models of DE. Thirdly, the microphysical origin of the cosine potential is of course different in the superfluid context. The potential depends on the difference of the phases of the superfluid, which are low-energy collective excitations superfluid, and it arises from a long range interaction within the entire superfluid. This is different than a self-interaction potential for a fundamental scalar field, as in the case of an axion.

This type of superfluid system is realized in Nature in known effects such the Josephson tunneling~\cite{Josephson,Mahan}. It is observed in laboratory experiments in models like Iron-based superconductors~\cite{Fernandes}, MgB2~\cite{Nicol}, high-$T_c$ cuprate superconductors or the $XY$ model~\cite{Schmitt:2014eka}. This gives us a good motivation for searching for an analogue of these models in the universe. The other advantage is that this model is minimal, only requiring the presence of this DM superfluid, providing a dynamical explanation for the cosmic acceleration without the presence of DE. This modification of the dynamics is obtained without modifying the underlying fundamental gravity theory of our universe.  So, we can say that this model unifies the DM and DE paradigms (and MOND), since we have an universe that has only DM, and where DE or the accelerated expansion is an emergent characteristic when considering a two-state superfluid. 

The idea of unified DM and DE is not new, and is commonly known in the literature as Unified Dark Matter (UDM), Unified Dark Energy (UDE), or quartessence~\cite{Makler:2002jv}. It comprises many models like the generalized Chaplygin gas~\cite{Bento:2002ps,Carturan:2002si,Sandvik:2002jz}, k-essence~\cite{Scherrer:2004au,Giannakis:2005kr}, and fast transition models~\cite{Bruni:2012sn,Leanizbarrutia:2017afj} (see~\cite{Bertacca:2010ct} for a extensive list of models), including also models that use condensates in this context~\cite{Cadoni:2017evg,Cadoni:2018dnd,Alexander:2009uu}.  These models usually have a stress-tensor that describes the observational eras, mimicking the background evolution from the matter-dominated and DE-dominated eras. Another recent example is~\cite{Berezhiani:2016dne}, where cosmic acceleration arises from suitable interactions between DM and baryons, without sources of negative pressure or new degrees of freedom beyond DM and ordinary matter.

However it has proven challenging to build a successful model that gives the desired background evolution together with a realistic and well-defined growth of perturbations, for the following reasons:

\begin{itemize}

\item In Chaplygin gas models, which postulates an exotic equation of state $P = -\rho^{-\alpha}$ for the dark fluid, the adiabatic sound speed becomes relativistic at late times, resulting either in large oscillations or late-time growth in the matter power spectrum~\cite{Sandvik:2002jz,Bertacca:2010ct}. This can be avoided if $\alpha$ assumes unnaturally small values, less than $10^{-5}$, in which case the model is indistinguishable from $\Lambda$CDM~\cite{Sandvik:2002jz}.

\item In ghost condensation~\cite{ArkaniHamed:2003uy} (and the closely related $k$-essence examples~\cite{Scherrer:2004au,Giannakis:2005kr}), the background energy density at the ghost condensate point mimics a cosmological constant and acts as DE, while small deviations in the scalar field away from the ghost condensate point behave as a pressureless fluid. Thanks to the spontaneous breaking of Lorentz invariance, linear perturbations have nearly vanishing sound speed, resulting in well-behaved perturbations in the linear regime. When these perturbations grow non-linear, however, the pressureless fluid develops caustics, whose resolution requires a UV completion for the ghost condensate~\cite{ArkaniHamed:2005gu}. Any attempt to describe DM as a fundamental scalar field with a non-trivial kinetic term, such as DBI models, faces the same issue~\cite{Frolov:2002rr,Felder:2002sv,DBIpaper,Khoury:2014tka}.
 
\end{itemize}

Our model naturally avoids both issues. Superfluid phonons are described by a non-linear kinetic term of the $P(X)$ form. The superfluid is assumed non-relativistic, hence its sound speed is suitably small throughout the evolution, which is given completely by the dynamics of the system, from the matter-dominated to the DE-dominated phase. At this level the situation is similar to ghost condensation (or, more precisely, tilted ghost condensation~\cite{Senatore:2004rj,Creminelli:2008wc}). A key difference, as mentioned earlier, is that the phonon field is not a fundamental scalar field. When perturbations become non-linear, the fluid caustics are naturally resolved --- the large gradients result in a local breakdown of superfluidity, the $P(X)$ effective description is no longer valid, and the physics is instead described by individual DM particles in the normal phase. This is the mundane ``UV completion" of the $P(X)$ phonon effective theory. (Once the halo virializes and achieves a quasi-static state, the superfluid phase is restored in the central region where DM thermalization and condensation occur.)

After this work was completed we became aware of~\cite{Braden:2017add}, where the authors consider two superfluid components with contact interactions. They arrive in a model that is similar to ours, with a potential like (\ref{cosinepotential_intro}) and extra terms giving then a sine Gordon theory for the angular degree of freedom. Their focus is to use this model as an analogue of false vacuum decay in quantum field theory, whereas we are interested in late-time cosmic acceleration. Furthermore, a key difference with our approach is that the superfluids describe DM.

\subsection{Outline}

This work is organized as follows. Section~\ref{SFDMreview} contains a review of the superfluid DM model. We begin by illustrating in Sec.~\ref{X2example} our mechanism in the simplest example of two weakly-coupled BEC DM superfluids. In Sec.~\ref{Sec: Constraints}, in particular, we use the results of the cosmological analysis to place constraints on the theory parameters, to ensure that the fluids' sound speeds are sufficiently small, and that the DE component approximates a cosmological constant, consistent with observations. 
We then generalize the analysis in Sec.~\ref{Sec. Model} and consider two coupled superfluids with arbitrary equations of state. 

In Sec.~\ref{cosmo} we turn our attention to the background cosmological evolution. Remarkably we will be able to derive a closed equation for the Hubble parameter that holds for any superfluid equation of state. We numerically integrate this background equation in Sec.~\ref{Sec. Numerics 1} for a set of parameter values that respect the various constraints derived in earlier Sections. 
In Sec.~\ref{Sec: Perturbations} we study the growth of linear perturbations in the model and derive an equation for the evolution of the total matter perturbation. This equation is numerically integrated 
in Sec.~\ref{pert num}, where we compare the growth function and growth rate for our model to the $\Lambda$CDM predictions. In a nutshell the growth function is close to $\Lambda$CDM, but the difference in growth rate is substantial at late times. We briefly summarize our results and mention future lines of research in Sec.~\ref{conclusions}. 

\section{Review of Superfluid DM}
\label{SFDMreview}

In order for DM to be in a superfluid phase in the central regions of galaxies, the particles must be light enough
that their de Broglie wavelengths overlap. This is a necessary condition for Bose-Einstein condensation, assuming weak coupling. In other words, the
de Broglie wavelength $\lambda_{\rm dB} \sim \frac{1}{mv}$ must be larger than the mean interparticle separation $\ell \sim (m/\rho)^{1/3}$. This translates
to an upper bound on the mass: $m \;\lsim\; (\rho/v^3)^{1/4}$.

As a back-of-the-envelope estimate, we can substitute $\rho_{200} \simeq 1.95 \times 10^{-27}~{\rm g}/{\rm cm}^3$, corresponding to 200 times the present critical density,
and $v = v_{200} \simeq  85~{\rm km}/{\rm s}$, corresponding to the velocity dispersion at the radius where the mean density for a Milky Way-like galaxy ($M = 10^{12}\,{\rm M}_\odot$)
reaches $\rho_{200}$. The result is
\beq
m \;\lsim\; 4.3~{\rm eV}\,.
\eeq
See~\cite{Berezhiani:2017tth} for a more refined version of this bound in the superfluid DM context using explicit density and velocity profiles. The general lesson is that superfluidity requires $m$ to be less than order eV. 

As a second requirement to form a superfluid, the DM particles should interact sufficiently
strongly  ($\sigma/m \gtrsim 0.1 \,\mathrm{cm}^2/\mathrm{g}$) to achieve
thermalization. Thus the particles are axion-like in the sense of being light and produced out of equilibrium, but because of the need for strong self-interactions
they cannot be QCD axions. 

As shown in~\cite{Berezhiani:2017tth}, the resulting density profile consists of a superfluid core, within which DM particles interact sufficiently frequently to thermalize and form a Bose Einstein condensate, surrounded by an envelope describing approximately collisionless particles following an NFW profile. The size of the superfluid region depends both on the model parameters as well as the total mass of the halo. 

In order for phonons to explain rotation curve properties, the superfluid core should at least encompass the entire range over which rotation curves are measured, {\it e.g.}, out to $\sim 60$~kpc for a Milky Way-like galaxy. Meanwhile, it is well known that galaxy clusters pose a problem for MOND. Therefore, in galaxy clusters, the superfluid region should be small enough that most of the DM is in the normal phase. In~\cite{Berezhiani:2017tth} it was shown that these requirements can be simultaneously satisfied for a range of parameters. For instance, with mass $m = 1~{\rm eV}$, cross-section $\frac{\sigma}{m} = 0.5~\frac{{\rm cm}^2}{\rm g}$ (satisfying the bound from merging clusters~\cite{Randall:2007ph,Massey:2010nd,Harvey:2015hha,Wittman:2017gxn}), and other fiducial parameter values reviewed below, the superfluid cores make up a modest fraction of the total DM mass in galaxies, ranging between $\sim 20\;\%$ for a $M \sim 10^{12} {\rm M}_\odot$ high-surface brightness galaxy to $\sim 50\;\%$ in a $M  \sim 10^{10} {\rm M}_\odot$ low-surface brightness galaxy. 

Therefore most of the mass is in the approximately collisionless NFW envelope, which has two advantages observationally: 1)~Halos are triaxial near the virial radius, exactly as in $\Lambda$CDM, and consistent with observations~\cite{Clampitt:2015wea}; 2)~Most of the gravitational lensing signal comes from the NFW envelope, thus no phonon-photon non-minimal coupling is required to reproduce galaxy-galaxy lensing statistics. Consequently both photons and gravitons propagate at the speed of light, consistent with GW170817~\cite{TheLIGOScientific:2017qsa}.

Within the superfluid region, the physics is described by a phase of a spontaneously broken global $U(1)$ symmetry. At low energy the relevant degrees of freedom are phonons, which are excitations
of the Goldstone boson $\theta$ for the broken symmetry. The effective theory of phonons must be invariant under the shift symmetry, $\theta \rightarrow \theta + c$, which non-linearly realizes the
$U(1)$ symmetry, and Galilean symmetry, appropriate for a non-relativistic superfluid. Therefore, its most general form at leading order in derivatives and zero temperature must be a
``$P(X)$" theory~\cite{Greiter:1989qb,Son:2002zn} 
\beq
{\cal L}_{\rm phonons} = P(X)\,;\qquad X= \dot{\theta} -m\Phi - \frac{(\vec{\nabla}\theta)^2}{2m}\,,
\label{LT=0}
\eeq
where $\Phi$ is the gravitational potential. The equation of state of the superfluid is encoded in the form of $P(X)$. Indeed, at finite chemical potential
$\theta = \mu t$, and with the gravitational potential set to zero, the Lagrangian reduces to $P(\mu)$, which defines the grand canonical equation of
state. In turn, the number density of particles in the condensate is $n = P_{,X}(\mu)$, with mass density $\rho = mn$. These
relations can be combined to deduce the pressure as a function of density, $P(\rho)$. The phonon sound speed is given by
$c_s^2=\frac{P_{,X}}{\rho_{,X}} = \frac{1}{m} \frac{P_{,X}}{P_{,XX}}$.

Superfluids are often described by a polytropic equation of state, $P(X) \sim X^n$, corresponding to $P(\rho) \sim \rho^{\frac{n}{n-1}}$. For instance, a standard, weakly-coupled superfluid corresponds to $n=2$:
\beq
P_{{\rm BEC}\;{\rm DM}}(X) = \frac{\Lambda^4}{2}\left(\frac{X}{m}\right)^2\,.
\label{BECDM}
\eeq
This is the case studied in the context of BEC DM~\cite{Sin:1992bg,Ji:1994xh,Goodman:2000tg,Peebles:2000yy,Arbey:2003sj,Boehmer:2007um,Lee:2008ux,Lee:2008jp,Harko:2011xw,Dwornik:2013fra,Guzman:2013rua,Harko:2014vya}. As another example, the Unitary Fermi Gas, describing a gas of ultra-cold fermionic atoms tuned at unitary, corresponds to $n = 5/2$~\cite{Son:2005rv}:
\beq
P_{\rm UFG}(X) = c_0 m^4 \left(\frac{X}{m}\right)^{5/2}\,,
\eeq
where $c_0$ is a dimensionless constant.  
 
The DM superfluid considered in~\cite{Berezhiani:2015bqa,Berezhiani:2015pia} corresponds to $n = 3/2$. More precisely, to ensure that the action is well-defined for $X> 0$ (time-like profile) and $X < 0$ (space-like profile), the actual superfluid action is
\beq 
P_{\rm DM}(X) =  \frac{2\Lambda (2m)^{3/2}}{3} X\sqrt{|X|}\,.
\label{PDM}
\eeq
The square-root form also ensures that the Hamiltonian is bounded from below~\cite{Bruneton:2007si}.  Modulo the square root and the fact that $\theta$ is a non-relativistic field,~\eqref{PDM} closely resembles the Bekenstein-Milgrom action to describe MONDian dynamics~\cite{Bekenstein:1984tv}. Importantly, unlike Bekenstein-Milgrom the additional long-range force
is not fundamental. Instead it is an emergent phenomenon due to collective excitations of DM. 

To mediate a force between baryons, the DM phonons must couple to the baryon density as
\beq
{\cal L}_{\rm int} = \alpha\Lambda\frac{\theta}{M_{\rm Pl}} \rho_{\rm b}\,,
\label{coupling}
\eeq
where $\alpha$ is a constant. (The reduced Planck scale $M_{\rm Pl}$ is related to the gravitational constant by $M_{\rm Pl}^2 = \frac{1}{8\pi G_{\rm N}}$.) This coupling explicitly breaks the $U(1)$ shift symmetry, albeit softly given the $M_{\rm Pl}$ suppression, making $\theta$ a pseudo-Goldstone boson. Physically, this operator implies that DM particle number conservation is violated in the presence of baryons. Its condensed matter origin is unclear
and begs for a compelling interpretation. (Alternatively, the MOND-like behavior can come from next-to-leading order terms in the derivative expansion for phonons~\cite{Khoury:2016ehj}.) 

In the regime where phonon gradients dominate\footnote{The phonon effective field theory breaks down for large phonon gradients, like in the vicinity of stars (e.g. in our solar system). For a more detailed description of the validity of the effective theory see Sec.~5 of~\cite{Berezhiani:2015bqa}.}, such that $X\simeq - \frac{(\vec{\nabla}\theta)^2}{2m}$, the phonon-mediated acceleration matches the deep-MOND expression
\beq
a_{\rm phonon} = \sqrt{a_0 a_{\rm b}}\,,
\eeq
where $a_{\rm b}$ is the Newtonian gravitational acceleration due to baryons only. The critical acceleration $a_0$ is related to the theory parameters as
\beq
a_0 = \frac{\alpha^3\Lambda^2}{M_{\rm Pl}}\,.
\label{a0us}
\eeq
The total force experienced by baryons is the sum of the phonon-mediated force and the Newtonian gravitational acceleration due to baryons and the DM condensate itself. 

Using a combination of analytical and numerical calculations,~\cite{Berezhiani:2017tth} showed that explicit rotation curves of both high and low surface brightness galaxies could
be reproduced in the superfluid model, with excellent agreement with observations. Interestingly, in contrast with MOND where the rotation curve becomes flat,~\cite{Berezhiani:2017tth} 
found a slight rise in the asymptotic velocity of massive galaxies. This is due to the gravitational pull of the superfluid DM mass itself, which can be as large as $\sim 30\%$ of the total force.
To compensate for this effect,~\cite{Berezhiani:2017tth} found a best-fit value for the critical acceleration of
\beq
a_0 \simeq 0.87\times 10^{-8}~{\rm cm/s^2}\,,
\eeq
which is somewhat lower than the MOND best-fit value of $1.2 \times 10^{-8}~{\rm cm}/{\rm s}^2$. For the record, the corresponding parameter values were $\Lambda = 0.05~{\rm meV}$ and $\alpha = 5.7$.

We have glossed over a subtlety of the superfluid framework, namely that perturbations around the static ``MONDian" background are unstable (ghost-like). As argued in~\cite{Berezhiani:2015bqa,Berezhiani:2015pia}, this instability can naturally be cured by finite-temperature corrections, which are expected given the non-zero temperature ({\it i.e.}, velocity dispersion) of DM particles in galactic halos. The temperature dependence in the effective theory is required independently to obtain an acceptable background cosmology and linear perturbation growth. To simplify the present analysis, we will ignore the issue of temperature dependence and focus our attention on the zero-temperature $P(X)$ Lagrangian.

\section{A Simple Example: Weakly-Coupled Non-Degenerate Superfluids} \label{X2example}

In the previous section we showed that a single species superfluid can describe the DM of our universe describing the MOND and particle-like dynamics of DM on small and large scales. In this section, we are going to generalize the model to a model where DM consists of two non-relativistic superfluid species, described in terms of two distinct phonon excitations. For instance, these could represent two distinguishable states of DM with slightly different energies, $\Delta E \ll m$, such as a ground state and an excited state. As shown in the Appendix, our two-component superfluid can be described in the mean-field approximation by two complex scalar fields, $\Psi_i=\frac{\rho_i}{\sqrt{2}} e^{{\rm i} \Theta_i}$, $i=1,2$.

To set the stage, it is instructive to consider the simplest example of two weakly-coupled BEC DM non-relativistic superfluids~\eqref{BECDM}, with mass $m_1$ and $m_2$, respectively:
\beq
\mathcal{L} =  \frac{\Lambda_1^4}{2m_1^2} X_1^2 +\frac{\Lambda_2^4}{2m_2^2} X_2^2\,;\qquad X_i \equiv \dot{\theta}_i -m_i\Phi - \frac{(\vec{\nabla}\theta_i)^2}{2m_i}\,,
\label{Ltheta1theta2_starting_point}
\eeq
where $\theta_i$ is the phonon excitation for the $i^{\mathrm{th}}$ field, coming from $\Theta_i=m_i t+\theta_i$. In the Appendix we review how the above theory derives as the non-relativistic limit of a Lorentz-invariant theory of two complex scalar fields with quartic interactions. Thus far the two species are non-interacting, and the above theory enjoys a $U\left(1\right)\times U\left(1\right)$ global symmetry, describing particle number conservation of each species separately. The conserved number densities are
\beq
n_i = \Lambda_i^4 \frac{X_i}{m_i^2} \,,
\label{n_i non-rel}
\eeq
The global $U(1)$ symmetries act non-linearly on the Goldstones as shift symmetries $\theta_i \rightarrow \theta_i + c_i$. 

We supplement the above theory with an interaction term, which breaks the symmetry group to a residual global symmetry
$U\left(1\right)\times U\left(1\right)\rightarrow U\left(1\right)$:
\begin{equation}
\mathcal{L}_{\rm int} \propto -\frac{\Psi_1^{*}\Psi_2+\Psi_2^{*}\Psi_1}{\left|\Psi_1\right|\left|\Psi_2\right|}\,.
\label{LintPsi1Psi2}
\end{equation}
This term is similar to a Rabi coupling in the mean-field approximation, except for the denominator, which is non-standard. This is important in order for the potential to have nearly constant magnitude and drive cosmic acceleration. As shown in the Appendix, in the non-relativistic regime this interaction between the different states results in an oscillatory potential that couples the two species together:
\beq
V(\theta_2-\theta_1 + \Delta E \,t) = M^4\cos^2 \left(\frac{\theta_2-\theta_1+ \Delta E \,t}{2}\right)\,,
\label{nonrelpotential}
\eeq
where $\Delta E \equiv m_2 - m_1$ is the energy difference between the two states.\footnote{The explicit time dependence in the potential can be understood from the relativistic description as arising from the background $\Theta_i=m_it+\theta_i$, with a time dependent term coming from the conserved charge density. Since the phonon fields transform non-linearly as $\theta_i \rightarrow - m_i c$ under time translations $t\rightarrow t + c$, the potential~\eqref{nonrelpotential} is Galilean invariant.} In other words, one can either think of the potential~\eqref{nonrelpotential} as being added directly in the non-relativistic theory of the phonon fields $\theta_1$ and $\theta_2$, or as resulting from the interaction term~\eqref{LintPsi1Psi2} in the mean-field description. In addition to the parameters from the DM superfluid model, $\Lambda_i$'s and $m_i$'s, the potential introduces the scale $M$, which sets the amplitude of the potential. To generate late-time cosmic acceleration, this will be set to the standard value $M \sim {\rm meV}$.

As mentioned in the Introduction, a cosine interaction is ubiquitous in systems with multiple superfluid/superconducting species,
interacting through Josephson or Rabi couplings. Consistent with the non-relativistic approximation, we assume the mass splitting is small compared to the mass, 
\beq
\Delta E \ll m_i\,.
\label{DelEwindow}
\eeq
Moreover, to simplify the analysis we assume that the mass splitting is large compared to $\dot{\theta}_2-\dot{\theta}_1$. We will verify {\it a posteriori} the validity of this assumption. In this case, the potential can be approximated as
\beq
V(\theta_2-\theta_1 + \Delta E \,t) \simeq V(\Delta E \,t) = M^4\cos^2 \left(\frac{\Delta E \,t}{2}\right)\,.
\label{NRcosine}
\eeq

The potential explicitly breaks the individual $U(1)$ symmetries down to the diagonal $U(1)$
subgroup that shifts the Goldstones by the same constant: $\theta_i \rightarrow \theta_i + c$. The charge densities $n_i$ are no longer separately
conserved, but the total density, 
\beq
n = \Lambda_1^4 \frac{X_1}{m_1^2} + \Lambda_2^4 \frac{X_2}{m_2^2} \,,
\label{mPX2}
\eeq
is conserved. This represents the total number density of DM particles in the condensate. 
Our effective description is valid as long as the number density of particles
in each condensate is positive,
\beq
n_i  \geq 0\,,
\label{positive_number_density_1}
\eeq
that is, as long as $X_i \geq 0$.

Adding the potential to the action, we obtain
\beq
\mathcal{L} =  \frac{\Lambda_1^4}{2m_1^2} X_1^2 +\frac{\Lambda_2^4}{2m_2^2} X_2^2  - (1-2\Phi) V(\theta_2-\theta_1 + \Delta E \,t)\,.
\label{Ltheta1theta2}
\eeq
Note that the coupling of $V$ to the gravitational potential follows in the weak-field limit by integrating out the spatial scalar potential $\Psi$ and ignoring $M_{\rm Pl}$-suppressed non-local terms. 
This action should be supplemented by the gravitational action ${\cal L}_{\rm grav} = - M_{\rm Pl}^2(\vec{\nabla}\Phi)^2$. 

For later purposes it will be convenient to record the sound speed of perturbations. Expanding the
kinetic term in~\eqref{Ltheta1theta2} around a time-dependent background $\bar{\theta}_i(t)$ to
quadratic order in field perturbations $\vartheta_i = \theta_i - \bar{\theta}_i(t)$, we obtain
\beq
{\cal L}_{\rm kin} = \sum_{i = 1}^2 \frac{\Lambda_i^4}{2m_i^2} \left(\dot{\vartheta}_i^2 - c_{s\, i}^2 \left(\vec{\nabla}\vartheta_i\right)^2\right)\,,
\label{LpertX2}
\eeq
where the sound speed of each component is
\beq
c_{s\, i}^2 = \frac{X_i}{m_i}  = \frac{m_i n_i}{\Lambda_i^4}\,.
\label{cs2}
\eeq
We can already see that for each of the superfluid DM species, the sound speed is always small given the non-relativistic approximation $X_i \ll m_i$, which tells us that these components cluster like the usual dust component. It is important to notice that this is a dynamical statement and it does not depend on the parameters of the model. 

\subsection{Diagonalization}

We can gain further intuition by performing the field redefinition
\beq
\xi  =  \frac{1}{\sqrt{\frac{\Lambda_1^4}{m_1^2} + \frac{\Lambda_2^4}{m_2^2}}} \left(\frac{\Lambda_1^4}{m_1^2}\theta_1 + \frac{\Lambda_2^4}{m_2^2}\theta_2\right) \;;\qquad
\chi = \frac{1}{\sqrt{\frac{\Lambda_1^4}{m_1^2} + \frac{\Lambda_2^4}{m_2^2}}} \frac{\Lambda_1^2\Lambda_2^2}{m_1m_2}\left(\theta_2 - \theta_1\right)\,.
\label{diagfieldredef}
\eeq
This diagonalizes the kinetic term, ${\cal L} = \frac{1}{2} \dot{\xi}^2 + \frac{1}{2} \dot{\chi}^2 + \ldots$, but the spatial gradients are still mixed. For our purposes it will suffice to work at leading order in $\frac{\Delta E}{m_i}\ll 1$. The relevant terms in this approximation are
\begin{eqnarray}
\nonumber
\mathcal{L} &=& \frac{1}{2} \dot{\xi}^2 + \frac{1}{2} \dot{\chi}^2 - \left(\Lambda^2 \dot{\xi} + \frac{\Lambda_1^2\Lambda_2^2}{\Lambda^2} \frac{\Delta E}{m} \dot{\chi}\right)\Phi - \frac{\dot{\xi}(\vec{\nabla}\xi)^2}{2\Lambda^2}  \\
& &  -~\frac{\dot{\chi}}{\Lambda^2} \left(\vec{\nabla}\xi\cdot \vec{\nabla}\chi  - \frac{\Lambda_1^2\Lambda_2^2}{2\Lambda^4} \frac{\Delta E}{m} (\vec{\nabla}\xi)^2\right) - (1-2\Phi) V\left(\Delta E\,t + \frac{\Lambda^2m}{\Lambda_1^2\Lambda_2^2} \chi\right)\,,
\label{L2diag} 
\end{eqnarray}
where we have defined
\beq
m \equiv \frac{m_1 + m_2}{2}\,;\qquad \Lambda^2 \equiv \sqrt{\Lambda_1^4 + \Lambda_2^4}\,.
\label{mLambda}
\eeq
The last gradient term in~\eqref{L2diag}, proportional to $\frac{\Delta E}{m} (\vec{\nabla}\xi)^2$, appears naively suppressed, but as we will
see it makes an unsuppressed contribution to the hydrodynamical equations and hence must be kept.

In this form, we recognize a Goldstone boson $\xi$ associated with total particle number conservation, and a pseudo-Goldstone boson $\chi$ with periodic potential:
\beq
V = M^4\cos^2 \left(\frac{\Delta E \,t}{2} + \frac{\Lambda^2m}{\Lambda_1^2\Lambda_2^2}  \frac{\chi}{2} \right)\,.
\eeq
Since both fields have canonical kinetic terms, we can immediately identify the physical decay constant in the cosine potential as
\beq
f_\chi = \frac{\Lambda_1^2\Lambda_2^2}{\Lambda^2m}\,.
\label{fchi}
\eeq
We will see that, unlike pNGB models of DE, $f_\chi$ is not forced to assume super-Planckian values in order for the 
potential to drive approximate de Sitter expansion. Meanwhile, expanding $V$ to quadratic order in $\chi$, the effective
mass for the pseudo-Goldstone boson is readily identified:
\beq
m_\chi^2 = \frac{M^4}{f_\chi^2} \cos \left(\frac{\Delta E \,t}{2}\right) \,.
\label{mchi}
\eeq

\subsection{Hydrodynamical equations}

Since our theory describes two interacting superfluids, it is instructive to write down their equations of motion in terms of fluid variables. 
The continuity and Euler's equations are first-order equations, hence to derive them we must work in the Hamiltonian description.

The conjugate phonon momenta follow immediately from~\eqref{L2diag}:
\begin{eqnarray}
\nonumber
\Pi_\xi &=& \frac{\partial{\cal L}}{\partial \dot{\xi}} = \dot{\xi} - \Lambda^2 \Phi - \frac{1}{2\Lambda^2} (\vec{\nabla}\xi)^2\,; \\
\Pi_\chi &=& \frac{\partial{\cal L}}{\partial \dot{\chi}}  =\dot{\chi} - \frac{\Lambda_1^2\Lambda_2^2}{\Lambda^2} \frac{\Delta E}{m} \Phi - \frac{1}{\Lambda^2} \left(\vec{\nabla}\xi\cdot \vec{\nabla}\chi  - \frac{\Lambda_1^2\Lambda_2^2}{2\Lambda^4} \frac{\Delta E}{m} (\vec{\nabla}\xi)^2\right)\,.
\label{Pi's}
\end{eqnarray}  
The Hamiltonian density ${\cal H} = \dot{\xi}\Pi_\xi + \dot{\chi}\Pi_\chi - {\cal L}$ is then 
\begin{eqnarray}
\nonumber
{\cal H} &=& \frac{1}{2} \Pi_\xi^2 + \frac{1}{2} \Pi_\chi^2 + \left(\Lambda^2 \Pi_\xi +  \frac{\Lambda_1^2\Lambda_2^2}{\Lambda^2} \frac{\Delta E}{m} \Pi_\chi \right)\Phi \\
& & +~\frac{\Pi_\xi}{2\Lambda^2}(\vec{\nabla}\xi)^2 + \frac{\Pi_\chi}{\Lambda^2} \left(\vec{\nabla}\xi\cdot \vec{\nabla}\chi  - \frac{\Lambda_1^2\Lambda_2^2}{2\Lambda^4} \frac{\Delta E}{m} (\vec{\nabla}\xi)^2\right) + (1-2\Phi) V\left(\Delta E\,t + \frac{\Lambda^2m}{\Lambda_1^2\Lambda_2^2} \chi\right)\,.
\label{hamiltonian}
\end{eqnarray}  
Hamilton's equations of motion are
\begin{eqnarray}
\nonumber
\dot{\xi} = \frac{\partial{\cal H}}{\partial\Pi_\xi} &=&  \Pi_\xi + \Lambda^2 \Phi + \frac{1}{2\Lambda^2} (\vec{\nabla}\xi)^2\,; \\
\nonumber
\dot{\chi} = \frac{\partial{\cal H}}{\partial\Pi_\chi} &=& \Pi_\chi + \frac{\Lambda_1^2\Lambda_2^2}{\Lambda^2} \frac{\Delta E}{m} \Phi + \frac{1}{\Lambda^2} \left(\vec{\nabla}\xi\cdot \vec{\nabla}\chi  -  \frac{\Lambda_1^2\Lambda_2^2}{2\Lambda^4} \frac{\Delta E}{m} (\vec{\nabla}\xi)^2\right)\,;\\ 
\nonumber
\dot{\Pi}_\xi = - \frac{\partial{\cal H}}{\partial\xi} &=& \frac{1}{\Lambda^2} \vec{\nabla}\cdot\left[\Pi_\xi \vec{\nabla}\xi +  \Pi_\chi \left(\vec{\nabla}\chi - \frac{\Lambda_1^2\Lambda_2^2}{\Lambda^4} \frac{\Delta E}{m} \vec{\nabla}\xi\right)\right]\,;\\
\dot{\Pi}_\chi = - \frac{\partial{\cal H}}{\partial\chi} &\simeq& \frac{1}{\Lambda^2} \vec{\nabla}\cdot\left(\Pi_\chi \vec{\nabla}\xi \right) - \frac{\Lambda^2m}{\Lambda_1^2\Lambda_2^2} V'(\Delta E\,t)\,,
\label{hamilton_eom}
\end{eqnarray}
where in the last step we have assumed, as in~\eqref{NRcosine}, that $V\left(\Delta E\,t + \frac{\Lambda^2m}{\Lambda_1^2\Lambda_2^2} \chi\right) \simeq V\left(\Delta E\,t \right)$. Moreover, here and henceforth, we have defined 
\beq
V'(x) \equiv \frac{{\rm d}V(x)}{{\rm d}x}\,.
\label{Vprime def}
\eeq
In other words, $V'(\Delta E\,t) \equiv \frac{{\rm d}V(\Delta E\,t)}{{\rm d}(\Delta E\,t)}$.

To cast the above equations as fluid equations, we must identify the hydrodynamical variables. The fluid densities
$\rho_\xi$ and $\rho_\chi$ can be read off from the coefficient of $\Phi$:
\beq
\rho_\xi = \Lambda^2 \,\Pi_\xi\,;\qquad \rho_\chi = \frac{\Lambda_1^2\Lambda_2^2}{\Lambda^2} \frac{\Delta E}{m} \,\Pi_\chi\,.
\label{rho xi rho chi def}
\eeq
Meanwhile, the fluid velocities can be identified by taking the spatial gradient of the first two equations in~\eqref{hamilton_eom}. By the Equivalence Principle, the result 
should be interpreted as $\dot{\vec{u}}_\xi = - \vec{\nabla}\Phi + \ldots$,  $\dot{\vec{u}}_\chi = - \vec{\nabla}\Phi + \ldots$ This allows us to identify:
\beq
\vec{u}_\xi = - \frac{\vec{\nabla}\xi}{\Lambda^2}\,;\qquad \vec{u}_\chi = -  \frac{\Lambda^2}{\Lambda_1^2\Lambda_2^2} \frac{m}{\Delta E}\vec{\nabla}\chi\,.
\eeq
In terms of these fluid variables, and the relative velocity $\vec{u} \equiv \vec{u}_\xi - \vec{u}_\chi$, the equations of motion~\eqref{hamilton_eom} become
\begin{eqnarray}
\nonumber
&& \dot{\vec{u}}_\xi  + \left(\vec{u}_\xi\cdot \vec{\nabla}\right) \vec{u}_\xi  = - \frac{1}{\Lambda^4} \vec{\nabla}\rho_\xi - \vec{\nabla}\Phi\,;\\
\nonumber
&& \dot{\vec{u}}_\chi  + \left(\vec{u}_\chi\cdot \vec{\nabla}\right) \vec{u}_\chi - \left(\vec{u}\cdot \vec{\nabla}\right) \vec{u} = - \frac{\Lambda^4}{\Lambda_1^4\Lambda_2^4}\left(\frac{m}{\Delta E}\right)^2 \vec{\nabla}\rho_\chi- \vec{\nabla}\Phi\,;\\
\nonumber
&& \dot{\rho}_\xi + \vec{\nabla}\cdot \left(\rho_\xi \vec{u}_\xi\right) - \vec{\nabla}\cdot \left(\rho_\chi \vec{u}\right) = 0\,;\\
& & \dot{\rho}_\chi + \vec{\nabla}\cdot \left(\rho_\chi \vec{u}_\xi\right) = - \Delta E \,V'\left(\Delta E\,t\right)\,.
\label{fluid_eoms}
\end{eqnarray}
The first pair of equations are Euler's equations; the second pair are continuity equations. Later on, in Sec.~\ref{Sec: Perturbations}, we will use the above hydrodynamical equations to derive the evolution of density perturbations.

\subsection{Phenomenological constraints} \label{Sec: Constraints}

We derive various constraints on the weakly-coupled model to ensure consistency with what is known observationally about DM and DE. The model involves four parameters $m_1$, $m_2$, $\Lambda_1$ and $\Lambda_2$, to describe the superfluid species, as well as two additional parameters $M$ and $f$ describing their interaction. As before we assume $m \sim {\rm eV}$ in order for DM to form a superfluid inside galaxies, and $M \sim {\rm meV}$ in order for the potential to drive cosmic acceleration at the present time. 

We now derive a constraint on the $\Lambda_i$'s, which later on will be used to constrain the decay constant $f_\chi$. To simplify the analysis, we assume that the parameters of the two superfluids are all of the same order. This is already true of the masses, $m_1\simeq m_2\simeq  m$, given the non-relativistic assumption $\Delta E\ll m_i$. But we further assume
\beq
\Lambda_1\simeq \Lambda_2\simeq \Lambda \,,
\label{allthesame}
\eeq
where $\Lambda$ was defined in~\eqref{mLambda}, as well as
\beq
\qquad n_1 \simeq n_2 \sim n\,.
\eeq
The assumption~\eqref{allthesame} is fairly natural if the two superfluids originate from different states of DM. 
It follows from~\eqref{cs2} that the sound speeds are nearly the same, and given by
\beq
c_{s\, i}^2 \sim \frac{m n}{\Lambda^4} \,.
\label{csequal}
\eeq
The number density is conserved, as mentioned earlier, which in a cosmological context implies that $\rho = m n \sim 1/a^3$.
Hence the sound speeds both redshift as
\beq
c_{s\, i}^2 \sim \frac{\rho_{\rm eq}}{2\Lambda^4} \frac{a_{\rm eq}^3}{a^3}\,,
\eeq
where the subscript``eq" denotes matter-radiation equality, and $\rho_{\rm eq} \simeq 0.4~{\rm eV}^4$ is the matter density at that time. 

An important constraint is that the fluids' sound speeds must be sufficiently small to satisfy observational constraints from the CMB and the large-scale structure.
The sound speed is important to determine the scale, the Jeans length, above which perturbations can grow through gravitational instability. Perturbations smaller than the Jeans length do not grow, and instead oscillate. In order to be consistent with the large scale structures we observe in our universe, the sound speed of our components must be small,  smaller than $10^{-6}$ at matter-radiation equality, so the Jeans length is sufficiently small that perturbations on scales of cosmological interest can grow unimpeded. 

To our knowledge, the constraints placed on the DM sound speed usually assume $c_s = {\rm const.}$, {\it e.g.},~\cite{Thomas:2016iav,Kopp:2018zxp}, with the result $c^2_{s} \;\lsim\; 10^{-6}$.
As such, this bound does not readily apply to our sound speeds, which rapidly decrease in time as $c_{s\, i}^2 \sim a^{-3}$. Nevertheless, as a highly conservative bound we will impose, at matter-radiation equality, 
\beq
c^2_{s,\,{\rm eq}} = \frac{\rho_{\rm eq}}{2\Lambda^4} \;\lsim\; 10^{-6}\,.
\label{cs eq bound}
\eeq
This is conservative in the sense that, once~\eqref{cs eq bound} is satisfied, then $c_s^2 \ll 10^{-6}$ at subsequent times. 
Using $\rho_{\rm eq} \simeq 0.4~\mathrm{eV}^4$, this implies a lower bound on the superfluid scale:
\beq
\Lambda \;\gsim\; 21~{\rm eV}\,.
\label{Lambdabound}
\eeq
Later on we will derive a far more stringent lower bound on $\Lambda$ --- see~\eqref{Lambdabound stringent}. In any case, the sound speeds remain suitably small throughout the epochs of matter domination and cosmic acceleration. This ensures that the growth of density perturbations proceeds uniformly on sub-horizon scales and avoids the undesirable features in the matter power spectrum seen in Chaplygin models~\cite{Sandvik:2002jz}.

\subsection{Self-consistency condition}

In various instances, such as~\eqref{NRcosine}, we will assume that $|\dot{\theta}_2-\dot{\theta}_1| \ll \Delta E$. We are now in a position to check the consistency of this approximation. Assuming as before that $\Lambda_1\simeq \Lambda_2\simeq \Lambda$ (and of course $m_1\simeq m_2\simeq  m$), the approximation in terms of $\chi$ amounts to
\beq
\Delta E \gg \frac{m}{\sqrt{2} \Lambda^2}  |\dot{\chi}|\,.
\label{Del E cond}
\eeq
On a cosmological background, the equation of motion for $\chi$ that follows from~\eqref{L2diag} is 
\beq
\ddot{\chi} + 3H\dot{\chi} \simeq \frac{M^4m}{2\Lambda^2} \sin \left(\Delta E\,t\right)\,.
\eeq
In the slow-roll approximation, we obtain $|\dot{\chi}|\;\lsim\; \frac{M^4m}{6\Lambda^2H}$, hence the condition~\eqref{Del E cond} reduces to
\beq
\Delta E \gg \frac{M^4m^2}{6\Lambda^4H}\,.
\eeq
This is most stringent at the present time, when $H = H_0$. Using $M^4 =3H_0^2M_{\rm Pl}^2$, we arrive at 
\beq
\frac{\Delta E}{H_0} \gg \frac{m^2M_{\rm Pl}^2}{2\Lambda^4}\,.
\label{Del E cond 2}
\eeq
We will use this inequality to derive a lower bound on $\Lambda$ in Sec.~\ref{acc phase}.

\section{General Superfluids} \label{Sec. Model}

The generalization of the non-relativistic action~\eqref{Ltheta1theta2} for arbitrary superfluids is straightforward:
\beq
\mathcal{L} = P_1(X_1) + P_2(X_2) - (1-2\Phi) V(\theta_2-\theta_1 + \Delta E \,t) \,.
\label{eq:model}
\eeq
Obtaining a general $P(X)$ theory from the Lorentz-invariant complex scalar field action is discussed in the Appendix.
We could of course consider adding derivative interactions between $X_1$ and $X_2$, but we focus on~\eqref{eq:model} for concreteness.\footnote{By adding interaction terms proportional to the derivatives, one could construct a kinetic term like $P(X_1,X_2)$. We do not do that in this paper, since we are only interested in interactions that have an explicit motivation.} We will keep the $P_i(X_i)$'s general, though our primary interest lies in the cases $P(X) \sim X^2$,  corresponding to BEC DM, and $P(X) \sim X\sqrt{|X|}$, corresponding to the DM superfluid theory of~\cite{Berezhiani:2015bqa,Berezhiani:2015pia}. A simplification in the latter case is that we ignore baryons in our analysis, hence the interaction term~\eqref{coupling} is irrelevant. The potential $V$ will be kept general as well, though we will be primarily interested in the cosine potential~\eqref{NRcosine}. As before, we will assume $V(\theta_1-\theta_2 + \Delta E \,t) \simeq V(\Delta E \,t)$ in the equations of motion.

The conserved charge is the total number density of DM particles in the superfluid state:
\beq
n \simeq P_{1\,,X_1} + P_{2\,,X_2}\,.
\label{totalcharge}
\eeq
As before, our effective description is valid as long as the number density of particles in each condensate is positive,
\beq
n_i \simeq P_{i\,,X_i} \geq 0\,.
\label{PXcond}
\eeq
In the non-relativistic approximation, the energy density of the superfluids is given by their rest mass energy plus the potential energy
\begin{eqnarray}
\nonumber
\rho &=&  m_1 P_{1\,,X_1} + m_2 P_{2\,,X_2} + V(\Delta E \,t) \\
&= & \frac{1}{2} (m_1 + m_2) n + \frac{1}{2} \Delta E \left(P_{1\,,X_1} - P_{2\,,X_2}\right) + V(\Delta E \,t)\,.
\label{rhogen}
\end{eqnarray}
Meanwhile, the pressure is given as usual by the Lagrangian density
\beq
{\cal P} = P_1(X_1) + P_2(X_2) - V(\Delta E \,t) \,.
\label{Pgen}
\eeq

The adiabatic sound speed of each species, governing the growth of perturbations, can be obtained once again by 
expanding~\eqref{eq:model} to quadratic order in field perturbations $\vartheta_i = \theta_i - \bar{\theta}_i(t)$. The result is
\beq
{\cal L}_{\rm kin} = \sum_{i = 1}^2 \frac{P_{i\,,X_i}}{2m_i c_{s\, i}^2} \left(\dot{\vartheta}_i^2 - c_{s\, i}^2 \left(\vec{\nabla}\vartheta_i\right)^2\right)\,,
\label{Lquadgen}
\eeq
with the sound speed, in the non-relativistic limit, given by
\beq
c_{s\, i}^2 = \frac{{\cal P}_{i\,,X_i}}{m_i n_{i\,,X_i}} = \frac{P_{i\,,X_i}}{m_i P_{i\,,X_iX_i}}\,.
\label{eq. cs}
\eeq
This agrees with~\eqref{cs2} in the particular case $P_i(X_i) = \frac{\Lambda_i^4}{2} \left( \frac{X_i^2}{m_i^2 }\right)$.
To ensure stability, $c_{s\, i}^2$ should be positive definite. In light of~\eqref{PXcond}, this requires
\beq
P_{i\,,X_i X_i} \geq 0\,.
\eeq
As we saw in Sec.~\ref{X2example}, the sound speed is always positive and small in the non-relativistic regime for the case where the kinetic term was quadratic. This is also automatically satisfied in the case where the non-standard kinetic term can be described by a power-law, $P(X) \propto \left(X/m \right)^n$, as seen in Sec.~\ref{SFDMreview}. The sound speed of each component in this case is given by $c_{s\, i}^2 \propto \left( X_i/m_i \right)$ which is positive and much smaller than unity in the non-relativistic regime $X_i \ll m_i$. This smallness of the sound speed is a dynamical statement in our theory, not depending on the parameters of the model, and is important so the perturbations of this model produce the large scales structures we observe in our universe. 

As before, it is helpful to translate the scale frequency of the cosine to a physical mass scale analogous to an axion decay constant. 
Similarly to the field redefinition~\eqref{diagfieldredef}, the pseudo-Goldstone $\chi$ is now given by the difference of perturbations:
\beq
\chi = \frac{\vartheta_2 - \vartheta_1}{\sqrt{\frac{m_1c_{s\, 1}^2}{P_{1\,,X_1}}+ \frac{m_2c_{s\, 2}^2}{P_{2\,,X_2}}}} \,.
\eeq
The physical decay constant in the cosine potential is readily identified as 
\beq
f_\chi = \frac{1}{\sqrt{\frac{m_1c_{s\, 1}^2}{P_{1\,,X_1}}+ \frac{m_2c_{s\, 2}^2}{P_{2\,,X_2}}}}\,.
\label{fchigen}
\eeq
The mass of $\chi$ is once again given by~\eqref{mchi}.

\section{Background evolution for general superfluids}
\label{cosmo}

We now turn to the study of the background cosmology. Remarkably we can derive a universal equation for the Hubble parameter,
valid for {\it any} superfluid equation of state. To see this, note that, using the total energy density~\eqref{rhogen}, the Friedmann equation
for a spatially-flat universe is
\beq
3H^2 M_{\rm Pl}^2 = \rho_+ + \rho_- + V(\Delta E \,t) \,,
\label{Friedmann}
\eeq
where we have defined
\beq
\rho_+ \equiv m n = m\left(P_{1\,,X_1} + P_{2\,,X_2}\right) \,;\qquad \rho_- \equiv \frac{1}{2} \Delta E \left(P_{1\,,X_1} - P_{2\,,X_2}\right)\,,
\eeq
with $m = \frac{1}{2}(m_1 + m_2)$. On the other hand, using~\eqref{rhogen} and~\eqref{Pgen}, the second Friedmann equation (the ``$\dot{H}$" equation) is
\beq
\dot{H}M_{\rm Pl}^2 = -\frac{1}{2} \left(\rho + {\cal P} \right) \simeq -\frac{1}{2}\left( \rho_+ + \rho_-\right)\,,
\label{dotH}
\eeq
where we have used the non-relativistic approximation $m_i P_{i\,,X_i}\gg P_i$. Combining these two equations, we obtain
\beq
\boxed{2\dot{H} + 3H^2 = \frac{V(\Delta E \,t)}{M_{\rm Pl}^2}\,.}
\label{Hclosed}
\eeq
As claimed, this is a closed equation for $H(t)$, which holds for any choice of $P_i(X_i)$. This equation is similar to what one finds in ghost condensation~\cite{ArkaniHamed:2003uy}.

Equation~\eqref{Hclosed} is all we need to solve for the background cosmology. For completeness it is also instructive to consider the phonon equations of motion:
\beq
\frac{{\rm d}}{{\rm d}t} \left(a^3P_{i\,,X_i} \right) \simeq (-1)^{i+1}a^3V'(\Delta E \,t)\,; \qquad i = 1,2\,,
\label{eq:our_EOM}
\eeq
where we have used the fact that $V$ only depends on the difference $\theta_2 - \theta_1$. As before, $V'$ denotes differentiation with respect to its argument --- see~\eqref{Vprime def}.
The sum of these two equations implies
\beq
\frac{{\rm d}}{{\rm d}t} \left(a^3\rho_+\right) = 0 \qquad \Longrightarrow \qquad \rho_+ \sim \frac{1}{a^3}\,.
\eeq
This is nothing but the conservation of the total number density of DM particles~\eqref{totalcharge}.
Meanwhile, the difference of the two equations~\eqref{eq:our_EOM} implies
\beq
\dot{\rho}_-+3H\rho_- = \Delta E\,V'(\Delta E \,t)\,.
\label{differenceeqn}
\eeq
This is similar to the equation for a canonical relativistic scalar, except that the potential is explicitly time-dependent.\footnote{By adding couplings that involve derivatives of the fields, one could instead obtain a theory with a non-canonical kinetic term for the field difference, {\it i.e.}, $P(X_1 - X_2)$.}

We now show that the background evolution of this model can describe the evolution of our universe from a matter-dominated phase to an accelerated phase at late times. We will solve analytically the background evolution only for each of the phases separately.  We will come back in Sec.~\ref{Sec: Constraints} and find exact numerical solutions to the background equations in the simplest of weakly-coupled superfluids.

In the following we show the solutions for the dust dominated epoch and the late-time accelerating phase.

\subsection{Dust-dominated phase} 

Let us consider the dust-dominated epoch, well before the onset of cosmic acceleration, when the potential energy is negligible ({\it i.e.}, when $H^2 \gg M^4/M_{\rm Pl}^2$):
\beq
2\dot{H} + 3H^2 \simeq 0\,.
\eeq
The solution is $H = \frac{2}{3t}$, describing a matter-dominated universe. In this regime~\eqref{eq:our_EOM} reduce to $\frac{{\rm d}}{{\rm d}t} \left(a^3P_{i\,,X_i} \right)\simeq 0$,
telling us that each species is separately conserved:
\beq
P_{i\,,X_i} \sim a^{-3}\,.
\label{dotthetahomog}
\eeq
Not surprisingly, the total phonon energy density~\eqref{rhogen} describes a pressureless fluid:
\beq
\rho  = \rho_+ + \rho_- = m_1 P_{1\,,X_1} + m_2 P_{2\,,X_2} \simeq \frac{\rho_{\rm eq}}{2} \frac{a_{\rm eq}^3}{a^3} \,.
\label{rhodust}
\eeq

The above pressureless behavior was derived to leading order in $\frac{X_i}{m_i}\ll 1$. Of course, more precisely our fluids do have a small pressure, and correspondingly a small adiabatic sound speed given by~\eqref{eq. cs}. We must make sure that the $c_{s\, i}^2$'s are sufficiently small to satisfy observational constraints from the CMB and the large-scale structure. In Sec.~\ref{Sec: Constraints} we will derive detailed constraints from observations on the theory parameters for a fiducial example, $P(X) \sim X^2$.

Incidentally, the scalar equations of motion~\eqref{eq:our_EOM}, ignoring the potential, boil down to
\beq
\dot{X}_i = - 3Hmc_{s\, i}^2 \,.
\eeq
With $c_{s\, i}^2  > 0$, we see that $X_i$ {\it decreases in time} during the dust-dominated phase, hence the non-relativistic approximation $X_i\ll m_i$ becomes increasingly accurate.

\subsection{Late-time accelerating phase}
\label{acc phase}

The dust-dominated phase ends once the potential energy becomes a significant contribution to the energy density at the present time. This occurs when $H \sim M^2/M_{\rm Pl}$. 
In order for the universe to experience an approximate de Sitter phase, the potential energy should be approximately constant on a Hubble time, that is,
\beq
\left\vert \frac{{\rm d}\ln V(\Delta E \,t)}{{\rm d}t}\right\vert \ll H_0\,.
\label{slowrollgen}
\eeq
For the cosine potential~\eqref{NRcosine} we have 
\beq
\left\vert \frac{{\rm d}\ln V(\Delta E \,t)}{{\rm d}t}\right\vert = \frac{\Delta E}{2}\left\vert \tan \Delta E \,t \right\vert \,.
\eeq
Thus, for typical values of $\Delta E\,t$ the condition~\eqref{slowrollgen} will be satisfied if
\beq
\frac{\Delta E}{2H_0} \ll 1\,.
\label{fcons}
\eeq
Thus the mass splitting must be smaller than $H_0$ to ensure slow-roll.

From~\eqref{Del E cond 2}, this implies lower bound $\Lambda$:
\beq 
\Lambda \gg \sqrt{M_{\rm Pl} m}\,.
\label{Lambdabound stringent}
\eeq
For $m = 1~{\rm eV}$, for instance, this requires $\Lambda \gg 5\times 10~{\rm TeV}$. For the simplest case of two weakly-coupled superfluids discussed in Sec.~\ref{X2example},
this translates into a bound on the decay constant $f_\chi$ given by~\eqref{fchi} (with $\Lambda_1 = \Lambda_2 = \Lambda$):
\beq
\frac{f_\chi}{M_{\rm Pl}} = \frac{\Lambda^2}{M_{\rm Pl}m} \gg 1\,.
\label{fchiconsnumeric}
\eeq
Thus, like pNGB models of DE, our model requires a super-Planckian decay constant.

\begin{figure}[htb]
\centering
\includegraphics[scale=0.75]{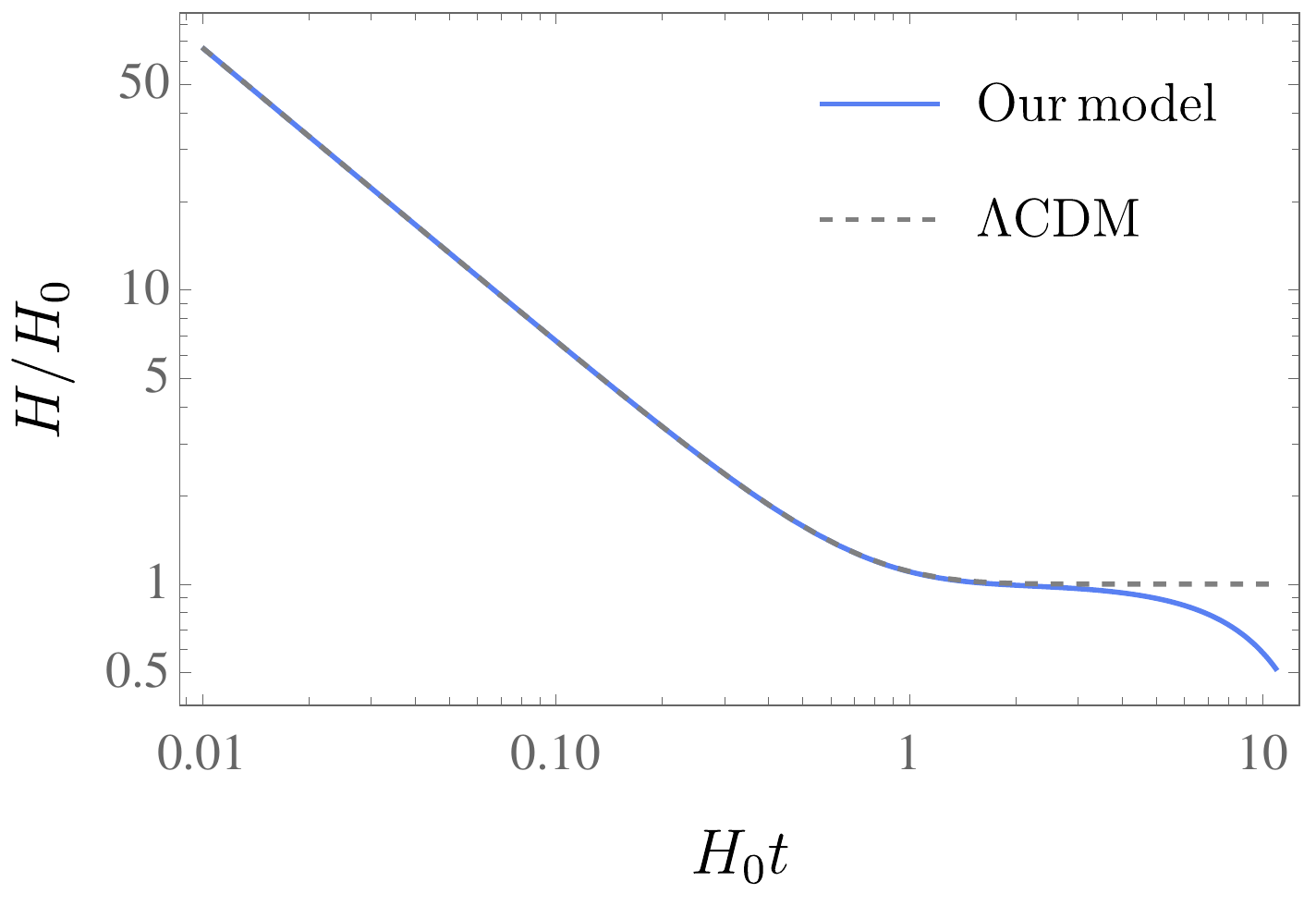}
\caption{Hubble parameter $H$ as a function of time for our model (blue solid curve) for the potential $V = M^4\cos^2 \left(\frac{\Delta E \,t}{2}\right)$ with $M = {\rm meV}$ and $\frac{\Delta E}{2H_0} = 0.1$, and a $\Lambda$CDM model (black dashed curve) with asymptotic Hubble constant $H_0^2 = M^4/3M_{\rm Pl}^2$.}
\label{fig:Hubble}
\end{figure}

\subsection{Numerical solution}
\label{Sec. Numerics 1}

To verify the above analytical arguments, we numerically solve the background cosmology of our model, for general superfluids. This amounts to integrating the evolution equation~\eqref{Hclosed} for the Hubble parameter. Since only the potential $V(\Delta E \,t)$ appears in that equation, we only need
to specify the potential parameters $M$ and $\Delta E$, subject to the constraints derived above:

\begin{itemize}

\item We set the scale of the potential to
\beq
M^4=3M_{\rm Pl}^2 H_0^2\sim {\rm meV}^4\,,
\eeq
with $H_0 \sim 10^{-33}\, \mathrm{eV}$ in order to derive cosmic acceleration today.

\item To satisfy the slow-roll condition~\eqref{fcons}, we set the energy gap, which is a characteristic of the microphysics of our superfluid, at
\beq
\frac{\Delta E}{2} = 0.2 \,H_0 \simeq 2\times 10^{-34}~{\rm eV}\,.
\eeq

\end{itemize}

Figure~\ref{fig:Hubble} shows the solution~\eqref{Hclosed} for the Hubble parameter for our model (blue solid curve) with the above parameters,
as well as a $\Lambda$CDM model (black dashed curve) with the same matter density at early times and asymptotic Hubble constant $H_0$. 
We can see that our model describes a phase of dust domination followed by a phase of accelerated expansion. The evolution is
indistinguishable from $\Lambda$CDM until the present time ($t_0 \sim H_0^{-1}$). Because our interaction potential is dynamical, the two expansion histories 
deviate from each other in the future, as the effects of the cosine potential come into play. The characteristic time at which the two histories deviate is set by the period of the cosine, {\it i.e.},
$\sim \frac{2H_0}{\Delta E} = 10\, H_0^{-1}$. Had we continued the evolution further in time, one would see the Hubble parameter
oscillate due to the cosine. However it is worth remembering that our superfluid effective theory is only valid in the regime when $\dot{H} \leq 0$, {\it i.e.},
whenever the Null Energy Condition is satisfied.

\section{Growth of Density Inhomogeneities} \label{Sec: Perturbations}

We have thus far showed that our model can describe the expansion history of our universe, with a period of decelerated evolution followed by a period of accelerated expansion, as in the $\Lambda$CDM model. A viable alternative to the $\Lambda$CDM model must not only reproduce the evolution of the background, but it should be able to describe the growth of density perturbations that leads to the structures we observe in our universe. In this section we turn to the analysis of density perturbations.

For simplicity we will focus on the interacting BEC DM superfluids of Sec.~\ref{X2example}. Our starting point is the set of hydrodynamical equations given by~\eqref{fluid_eoms}. Because these were derived in the weak-field approximation, they can be applied to the cosmological context in the freely-falling coordinate system of the Friedmann-Robertson-Walker (FRW) metric:
\beq
{\rm d}s^2 = -(1 + 2\Phi) {\rm d}t^2 + (1-2\Psi) {\rm d}\vec{\ell}^2\,.
\eeq
The proper distance $\ell$ is related to the comoving distance via $\vec{\ell} = a(t) \vec{x}$. The FRW background corresponds to 
\beq
\Phi_{\rm FRW} = -\frac{1}{2}\left(\dot{H} + H^2\right) \vec{\ell}^2\,;\qquad \Psi_{\rm FRW} = \frac{1}{4}H^2\vec{\ell}^2\,.
\eeq
This coordinate system is valid on sub-Hubble scales, $H \ell \ll 1$.

To make the assumed coordinate system explicit, let us rewrite the fluid equations~\eqref{fluid_eoms} as
\begin{eqnarray}
\nonumber
&& \left(\frac{\partial\rho_\xi}{\partial t}\right)_\ell  + \vec{\nabla}_\ell\cdot \left(\rho_\xi \vec{u}_\xi\right) - \vec{\nabla}_\ell \cdot \left(\rho_\chi \vec{u}\right) = 0\,;\\
\nonumber
&& \left(\frac{\partial\vec{u}_\xi}{\partial t}\right)_\ell  + \left(\vec{u}_\xi\cdot \vec{\nabla}_\ell \right) \vec{u}_\xi  = - \frac{1}{\Lambda^4} \vec{\nabla}_\ell\rho_\xi - \vec{\nabla}_\ell\Phi\,;\\
\nonumber
& &\left(\frac{\partial\rho_\chi}{\partial t}\right)_\ell+ \vec{\nabla}_\ell\cdot \left(\rho_\chi \vec{u}_\xi\right) = - \Delta E \,V'\left(\Delta E\,t\right)\,;\\
&& \left(\frac{\partial\vec{u}_\chi}{\partial t}\right)_\ell   + \left(\vec{u}_\chi\cdot \vec{\nabla}_\ell\right) \vec{u}_\chi - \left(\vec{u}\cdot \vec{\nabla}_\ell\right) \vec{u} = -  \frac{\Lambda^4}{\Lambda_1^4\Lambda_2^4}\left(\frac{m}{\Delta E}\right)^2 \vec{\nabla}_\ell \rho_\chi- \vec{\nabla}_\ell\Phi \,,
\label{fluid_eoms_FRW}
\end{eqnarray}
where $(\partial/\partial t)_\ell$ denotes time-differentiation at fixed $\vec{\ell}$. The gravitational potential is determined as usual by Poisson's equation:
\beq
\vec{\nabla}^2_\ell \Phi = \frac{1}{2M_{\rm Pl}^2} \left(\rho_\xi + \rho_\chi\right)\,.
\eeq

We now transform to the expanding coordinate system (see, {\it e.g.},~\cite{peeblesbook}), with $\vec{\ell} = a(t) \vec{x}$. The fluid densities are
\beq
\rho_\xi = \rho_\xi \left(\frac{\vec{\ell}}{a(t)}, t\right) \,;\qquad \rho_\chi = \rho_\chi \left(\frac{\vec{\ell}}{a(t)}, t\right) \,.
\eeq
The fluid velocities can be decomposed as
\beq
\vec{u}_\xi = H\vec{\ell} + \vec{v}_\xi \left(\frac{\vec{\ell}}{a(t)}, t\right)\,;\qquad \vec{u}_\chi = H\vec{\ell} + \vec{v}_\chi \left(\frac{\vec{\ell}}{a(t)}, t\right)\,.
\eeq
The first terms account for the Hubble flow, whereas $\vec{v}_\xi$, $\vec{v}_\chi$ are the peculiar velocities. The Hubble flow of course drops out of the relative velocity:
\beq
\vec{u} \equiv \vec{u}_\xi - \vec{u}_\chi = \vec{v}_\xi \left(\frac{\vec{\ell}}{a(t)}, t\right) - \vec{v}_\chi \left(\frac{\vec{\ell}}{a(t)}, t\right) \equiv \vec{v}\left(\frac{\vec{\ell}}{a(t)}, t\right) \,.
\eeq
Finally, the gravitational potential $\Phi$ can be split into a background piece and an inhomogeneous term:
\beq
\Phi = -\frac{1}{2}\left(\dot{H} + H^2\right) \vec{\ell}^{\;2} + \phi\left(\frac{\vec{\ell}}{a(t)}, t\right)\,.
\eeq
Making these substitutions, we obtain the hydrodynamical equations in the expanding coordinate system:
\begin{eqnarray}
\nonumber
&& \dot{\rho}_\xi + 3H\rho_\xi + \frac{1}{a}\vec{\nabla}\cdot \left(\rho_\xi \vec{v}_\xi\right) - \frac{1}{a}\vec{\nabla}\cdot \left(\rho_\chi \vec{v}\right) = 0\,;\\
\nonumber
&& \dot{\vec{v}}_\xi  + H \vec{v}_\xi +  \frac{1}{a} \left(\vec{v}_\xi\cdot \vec{\nabla}\right) \vec{v}_\xi  = - \frac{1}{\Lambda^4} \frac{\vec{\nabla}\rho_\xi}{a} - \frac{\vec{\nabla}\phi}{a}\,;\\
\nonumber
& & \dot{\rho}_\chi + 3H\rho_\chi  + \frac{1}{a}\vec{\nabla}\cdot \left(\rho_\chi \vec{v}_\xi\right) = - \Delta E \,V'\left(\Delta E\,t\right)\,; \\
&& \dot{\vec{v}}_\chi  + H \vec{v}_\chi + \frac{1}{a} \left(\vec{v}_\chi\cdot \vec{\nabla}\right) \vec{v}_\chi - \frac{1}{a}\left(\vec{v}\cdot \vec{\nabla}\right) \vec{v} = - \frac{\Lambda^4}{\Lambda_1^4\Lambda_2^4}\left(\frac{m}{\Delta E}\right)^2 \frac{\vec{\nabla}\rho_\chi}{a}- \frac{\vec{\nabla}\phi}{a}\,.
\label{fluid_eoms_expanding}
\end{eqnarray}

\subsection{Non-linear evolution of inhomogeneities}

Each fluid density can be decomposed into a background piece and an inhomogeneous term:
\beq
\rho_\xi = \bar{\rho}_\xi(t) + \delta\rho_\xi (\vec{x},t)\,;\qquad \rho_\chi = \bar{\rho}_\chi(t) + \delta\rho_\chi (\vec{x},t)\,.
\label{density_decomposition}
\eeq
Note that $\delta\rho_\xi$ and $\delta\rho_\chi$ are {\it not} assumed small at this stage. The background densities obey the equations
\begin{eqnarray}
\nonumber
&& \dot{\bar{\rho}}_\xi + 3H\bar{\rho}_\xi  = 0\,;\\
&&  \dot{\bar{\rho}}_\chi + 3H\bar{\rho}_\chi  = - \Delta E \,V'\left(\Delta E\,t\right)\,.
\label{background_eoms}
\end{eqnarray}
This confirms, in particular, that $\bar{\rho}_\xi$ describes dust and redshifts as $1/a^3$. Meanwhile, the evolution of $\rho_\chi$ is influenced by the potential.

It will be convenient to define perturbations relative to the {\it total} background density, $\bar{\rho} = \bar{\rho}_\xi + \bar{\rho}_\chi$,
\beq
\delta_\xi \equiv \frac{\delta\rho_\xi}{\bar{\rho}}\,;\qquad \delta_\chi \equiv \frac{\delta\rho_\chi}{\bar{\rho}}\,,
\label{del xi del chi new}
\eeq
where $\bar{\rho}$ satisfies
\beq
\dot{\bar{\rho}} + 3H\bar{\rho}  = - \Delta E \,V'\left(\Delta E\,t\right)\,.
\label{background eom total}
\eeq
Note that the total density perturbations is given by
\beq
\delta \equiv \frac{\delta\rho}{\bar{\rho}} = \delta_\xi + \delta_\chi\,.
\eeq

Substituting the decomposition~\eqref{density_decomposition} making use of the background equations~\eqref{background_eoms}, 
the fluid equations become
\begin{eqnarray}
\nonumber
&& \dot{\delta}_\xi + \frac{1}{a} \vec{\nabla}\cdot \left(\left(\frac{\bar{\rho}_\xi}{\bar{\rho}} + \delta_\xi\right) \vec{v}_\xi\right) - \frac{1}{a} \vec{\nabla}\cdot \left(\left(\frac{\bar{\rho}_\chi}{\bar{\rho}}+\delta_\chi\right) \vec{v}\right) =   \frac{\Delta E \,V'}{\bar{\rho}_\chi}\,\delta_\xi\,;\\
\nonumber
&& \dot{\vec{v}}_\xi  + H \vec{v}_\xi +  \frac{1}{a} \left(\vec{v}_\xi\cdot \vec{\nabla}\right) \vec{v}_\xi  = - \frac{\bar{\rho}}{\Lambda^4} \frac{\vec{\nabla}\delta_\xi}{a} - \frac{\vec{\nabla}\phi}{a}\,;\\
\nonumber
& & \dot{\delta}_\chi + \frac{1}{a} \vec{\nabla}\cdot  \left(\left(\frac{\bar{\rho}_\chi}{\bar{\rho}}+ \delta_\chi\right) \vec{v}_\xi\right)  = \frac{\Delta E \,V'}{\bar{\rho}_\chi}\,\delta_\chi\,; \\
&& \dot{\vec{v}}_\chi  + H \vec{v}_\chi + \frac{1}{a} \left(\vec{v}_\chi\cdot \vec{\nabla}\right) \vec{v}_\chi - \frac{1}{a}\left(\vec{v}\cdot \vec{\nabla}\right) \vec{v} = - \frac{\Lambda^4\bar{\rho}}{\Lambda_1^4\Lambda_2^4}\left(\frac{m}{\Delta E}\right)^2   \frac{\vec{\nabla}\delta_\chi}{a}- \frac{\vec{\nabla}\phi}{a}\,,
\label{fluid_eoms_inhomog}
\end{eqnarray}
together with Poisson's equation,
\beq
\vec{\nabla}^2 \phi = \frac{a^2}{2M_{\rm Pl}^2} \bar{\rho}\,\delta\,.
\label{poisson pert}
\eeq
The above set of equations are non-linear, Newtonian hydrodynamical equations in an expanding universe. They can be solved numerically to describe the
formation of non-linear structures in the Newtonian regime.

\subsection{Linear perturbations}

In the linear regime, where the $\delta$'s and $v$'s are all small,~\eqref{fluid_eoms_inhomog} simplify to
\begin{eqnarray}
\nonumber
&& \dot{\delta}_\xi + \frac{1}{a} \frac{\bar{\rho}_\xi}{\bar{\rho}}\vec{\nabla}\cdot \vec{v}_\xi -\frac{1}{a}\frac{\bar{\rho}_\chi}{\bar{\rho}} \vec{\nabla}\cdot \vec{v} =  \frac{\Delta E \,V'}{\bar{\rho}}\,\delta_\xi \,;\\
\nonumber
&& \dot{\vec{v}}_\xi  + H \vec{v}_\xi  = - \frac{\bar{\rho}}{\Lambda^4}  \frac{\vec{\nabla}\delta_\xi}{a} - \frac{\vec{\nabla}\phi}{a}\,;\\
\nonumber
&& \dot{\delta}_\chi + \frac{1}{a}  \frac{\bar{\rho}_\chi}{\bar{\rho}}\vec{\nabla}\cdot \vec{v}_\xi  = \frac{\Delta E \,V'}{\bar{\rho}}\,\delta_\chi\,; \\
&& \dot{\vec{v}}  + H \vec{v} =  -  \frac{\bar{\rho}}{\Lambda^4}  \frac{\vec{\nabla}\delta_\xi}{a} + \frac{\Lambda^4\bar{\rho}}{\Lambda_1^4\Lambda_2^4}\left(\frac{m}{\Delta E}\right)^2 \frac{\vec{\nabla}\delta_\chi}{a}\,.
\label{fluid_eoms_linear}
\end{eqnarray}
In particular, by adding the first and third equations we obtain an equation for $\delta = \delta_\xi + \delta_\chi$,
\beq
\dot{\delta} + \frac{1}{a} \vec{\nabla}\cdot \vec{v}_\xi - \frac{1}{a}\frac{\bar{\rho}_\chi}{\bar{\rho}} \vec{\nabla}\cdot \vec{v} =  \frac{\Delta E \,V'}{\bar{\rho}}\,\delta\,.
\eeq
The above equations, together with Poisson's equation~\eqref{poisson pert}, can be combined as usual to obtain a second-order equation for the density perturbation:
\begin{eqnarray}
\nonumber
\ddot{\delta} + \left(2H -  \frac{\Delta E \,V'}{\bar{\rho}}\right)\dot{\delta} &=&  \frac{1}{2M_{\rm Pl}^2} \bar{\rho}\,\delta - \frac{\bar{\rho}_\xi}{\bar{\rho}}\frac{\Delta E \,V'}{\bar{\rho}} \frac{\vec{\nabla}\cdot \vec{v}}{a} + c_{s,\,\xi}^2 \frac{\vec{\nabla}^2\delta_\xi}{a^2} + c_{s,\,\chi}^2 \frac{\vec{\nabla}^2\delta_\chi}{a^2} + \frac{\Delta E \,V'}{\bar{\rho}}\left(5H +\frac{\Delta E \,V'}{\bar{\rho}}\right) \delta\,;\\
\dot{\vec{v}}  + H \vec{v} &=&  - \frac{\bar{\rho}}{\bar{\rho}_\xi}c_{s,\,\xi}^2 \frac{\vec{\nabla}\delta_\xi}{a} + \frac{\bar{\rho}}{\bar{\rho}_\chi} c_{s,\,\chi}^2 \frac{\vec{\nabla}\delta_\chi}{a}\,,
\label{2nd_order_eoms}
\end{eqnarray}
where we have introduced the sound speed parameters:
\beq
c_{s,\,\xi}^2 \equiv \frac{\bar{\rho}_\xi}{\Lambda^4}\,;\qquad c_{s,\,\chi}^2 \equiv \frac{\Lambda^4\bar{\rho}_\chi}{\Lambda_1^4\Lambda_2^4}\left(\frac{m}{\Delta E}\right)^2 \,.
\label{sound_speeds}
\eeq
In deriving the first of~\eqref{2nd_order_eoms}, we have made use of $\frac{\Delta E \,V''}{V'} \sim \Delta E \ll H$, which follows from~\eqref{fchiconsnumeric}. 
By differentiating the $\delta$ equation once more, we could eliminate $\vec{v}$ using the second equation. In what follows we will instead argue that the spatial gradient terms can be ignored to obtain a single ordinary differential equation for the density perturbation.

\subsection{Ignoring spatial gradients}

The spatial gradient terms in~\eqref{2nd_order_eoms} are all proportional to the sound speed parameters $c_{s,\,\xi}^2$ and $c_{s,\,\chi}^2$. We now argue that these parameters are small, hence the spatial gradient terms can be neglected on linear scales. For $c_{s,\,\xi}^2$, this immediately follows from the analysis of Sec.~\ref{Sec: Constraints}, in particular~\eqref{cs eq bound} and~\eqref{Lambdabound}, since
\beq
c_{s,\,\xi}^2 = \frac{\bar{\rho}_\xi}{\Lambda^4} = \frac{mn}{\Lambda^4}\ll 1\,.
\eeq
For $c_{s,\,\chi}^2$, we shall assume for simplicity that $\Lambda_1\simeq \Lambda_2\simeq \Lambda$, as we did in~\eqref{allthesame}. Moreover, from~\eqref{background_eoms} we estimate that $\bar{\rho}_\chi \sim \frac{\Delta E \,V'}{H} \sim \frac{\Delta E \,M^4}{H}$. Thus we obtain
\beq
c_{s,\,\chi}^2 \sim \frac{m^2}{\Lambda^4} \frac{M^4}{H\Delta E}  \,.
\eeq
Unlike $c_{s,\,\xi}^2$, which redshifts as $1/a^3$, we see that $c_{s,\,\chi}^2$ grows as $1/H$ and therefore assumes its largest value at late times, {\it i.e.}, when $H = H_0$.
We have
\beq
c_{s,\,\chi}^2 \;\lsim\; \frac{m^2}{\Lambda^4} \frac{M^4}{H_0\Delta E} =\frac{m^2M_{\rm Pl}^2}{\Lambda^4} \frac{H_0}{\Delta E}\,.
\eeq
We have already required $m^2M_{\rm Pl}^2\ll \Lambda^4$ --- see~\eqref{Lambdabound stringent}. By imposing the slightly tighter bound $m^2M_{\rm Pl}^2\ll \Lambda^4 \frac{\Delta E}{H_0}$,
we ensure that $c_{s,\,\chi}^2 \ll 1$. Thus both modes can naturally have small sound speeds.

Ignoring spatial gradient terms, the first of~\eqref{2nd_order_eoms} reduces 
\beq
\boxed{\ddot{\delta} + \left(2H -  \frac{\Delta E \,V'}{\bar{\rho}}\right)\dot{\delta}  =  \frac{1}{2M_{\rm Pl}^2} \bar{\rho}\,\delta + \frac{\Delta E \,V'}{\bar{\rho}}\left(5H +\frac{\Delta E \,V'}{\bar{\rho}}\right) \delta\,.}
\label{2nd_order_eoms_final}
\eeq
Furthermore, the second of~\eqref{2nd_order_eoms} reduces to $\dot{\vec{v}}  + H \vec{v} \simeq 0$, which implies $\vec{v} \sim 1/a$. Thus the peculiar velocity difference redshifts away.

\section{Numerical Evolution of Density Perturbations}
\label{pert num}

 The second-order equation~\eqref{2nd_order_eoms_final} for governing the linear growth of density perturbations is straightforward to integrate numerically. 
For this purpose we assume the same parameter values as used in Sec.~\ref{Sec. Numerics 1}. We also set $m = 1~{\rm eV}$, and 
\beq
\Lambda_1 = \Lambda_2 = 500~{\rm eV}\,.
\eeq
This satisfies the bound~\eqref{Lambdabound} and ensures that the sound speeds are small, thereby justifying our neglecting the spatial gradients.  
Note that because we can drop the spatial gradients, the resulting equation of motion~\eqref{2nd_order_eoms_final} is independent of $\Lambda$. 
Furthermore, the decay constant~\eqref{fchi} evaluates to:
\beq
f_\chi = \frac{\Lambda^2}{m} \,f = 10^{25}~{\rm eV}\,.
\eeq
This is $\ll M_{\rm Pl}$, as claimed earlier. 

\subsection{Background solution for the matter densities}

The first step consists of solving~\eqref{background_eoms} for the background densities $\bar{\rho}_\xi$ and $\bar{\rho}_\chi$.
This requires specifying initial conditions, {\it i.e.}, the value of the densities at matter-radiation equality where we begin our integration.
Since our superfluid is a mixture of particles in their ground and excited states, the relative fraction of the two populations depends on
how the energy gap $\Delta E$ compares to the DM temperature at matter-radiation equality.
This in turn depends on the production mechanism of our DM particles.

Assuming that the production mechanism is an axion-like vacuum displacement, then by definition this will take place 
when $H \sim m = {\rm eV}$, corresponding to redshift $z_{\rm production} \sim 10^{16}$. Although the DM particles are initially
relativistic, $T_{\rm prod} \sim {\rm eV}$, soon thereafter they become non-relativistic, and their temperature subsequently redshifts as $1/a^2$. Hence by matter-radiation
equality ($z_{\rm eq} \sim 10^3$), the DM temperature has dropped to 
\beq
T_{\rm eq} \sim 10^{-26}~{\rm eV} \,.
\eeq
Thus, with our chosen value of $\Delta E = 5 \times 10^{-11}~{\rm eV}$, we have $T_{\rm eq} \ll \Delta E$. Hence, to a good approximation, all particles are in the ground state at that time. That is, the matter density is entirely in $\theta_1$.

This translates to the variables $\xi$ and $\chi$ in the following way. Using $\Lambda_1 = \Lambda_2$, we combine~\eqref{diagfieldredef},~\eqref{Pi's} and~\eqref{rho xi rho chi def}
to obtain $\bar{\rho}_\xi = \frac{\Lambda^4}{2m}\left(\dot{\bar{\theta}}_1+\dot{\bar{\theta}}_2\right)$ and $\bar{\rho}_\chi = \frac{\Lambda^4}{4m}\frac{\Delta E}{m} \left(\dot{\bar{\theta}}_2-\dot{\bar{\theta}}_1\right)$. Therefore, since we have just argued that $\dot{\bar{\theta}}_2 \ll \dot{\bar{\theta}}_1$ at matter-radiation equality, we obtain
\beq
\frac{\bar{\rho}_\chi^{\;\rm eq}}{\bar{\rho}_\xi^{\;\rm eq}} = -\frac{\Delta E}{2m} \,.
\label{rho chi eq}
\eeq
Since $\Delta E \ll m$, almost all of the mass density is in $\xi$ initially, hence 
\beq
\bar{\rho}_\xi^{\;\rm eq} \simeq \rho_{\rm eq} = 0.4~{\rm eV}^4\,.
\label{rho xi eq}
\eeq
Equations~\eqref{rho chi eq} and~\eqref{rho xi eq} give the initial conditions for the evolution of the energy densities. Note from~\eqref{rho chi eq} that $\rho_\chi$ is initially negative, however this is an artifact of the definition of the variables $\xi$ and $\chi$ in~\eqref{diagfieldredef}. It does not imply any violation of energy conditions, since, as seen in Secs.~\ref{X2example} and~\ref{Sec. Model}, the density of the phonon variables $\theta_1$ and $\theta_2$ is always positive.

\begin{figure}[htb]
\includegraphics[scale=0.7]{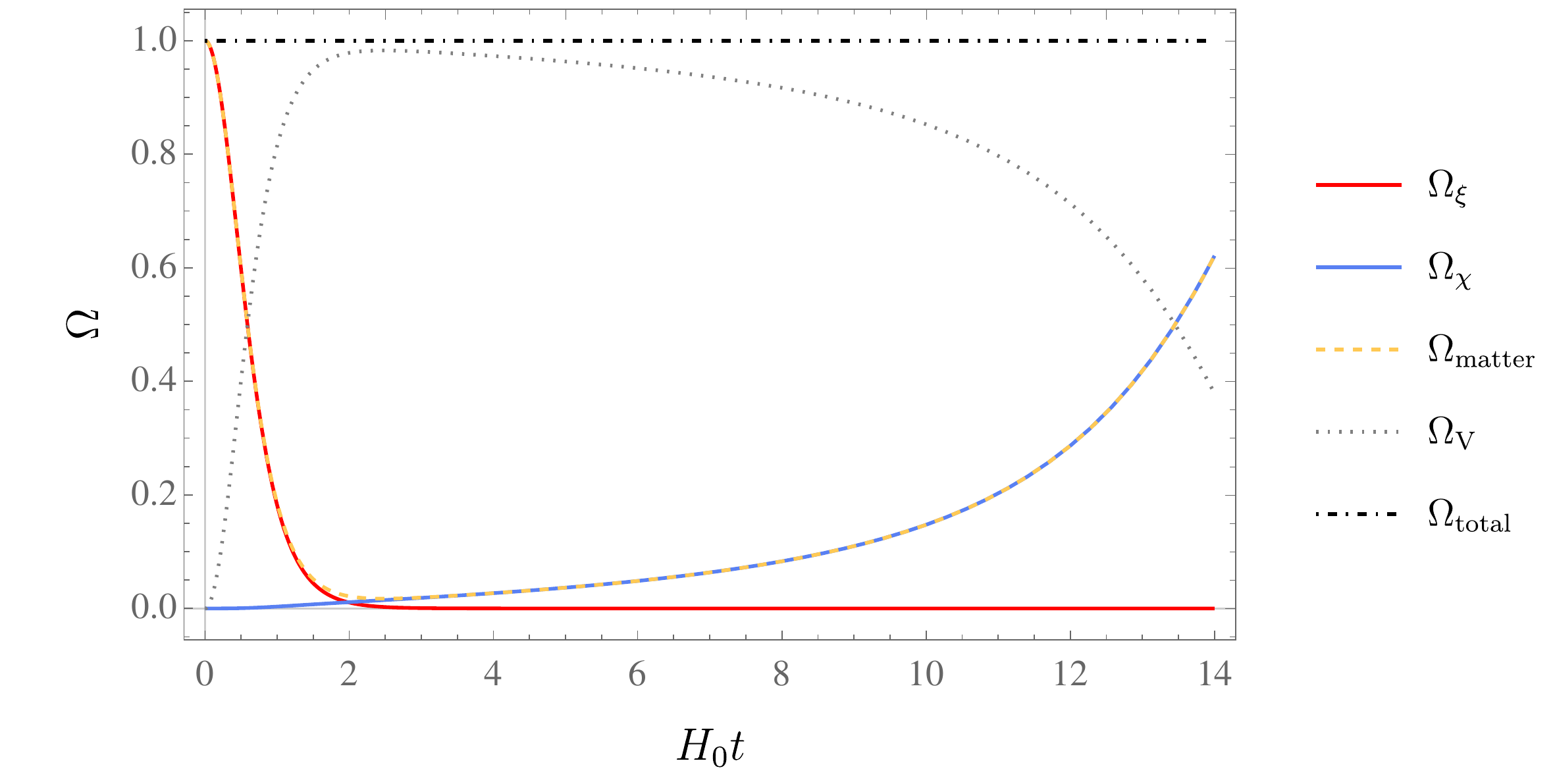}
\caption{Evolution of the fraction density parameters $\Omega_x = \frac{\rho_x}{3M_{\rm Pl}^2 H^2}$ for the different components: $\xi$ (red), $\chi$ (blue), the total matter density given by their sum (yellow), and the potential energy (dotted gray). The potential dominates close to the present time, resulting in cosmic acceleration. Due to the oscillatory nature of the potential, the matter component dominates again in the future.}
\label{fig:densities_future}
\end{figure}

Figure~\ref{fig:densities_future} shows the evolution, from matter-radiation equality onwards, of the fractional density parameters $\Omega_x = \frac{\rho_x}{3M_{\rm Pl}^2 H^2}$ for our different components: $\Omega_\xi$ (red), $\Omega_\chi$ (blue), the total matter density $\Omega_\xi + \Omega_\chi$ (yellow), and the potential energy $\Omega_V$ driving late-time accelerated expansion (dotted gray). Since we assume a spatially-flat universe, the $\Omega$'s add up to unity at all times, $\Omega_{\rm total} = 1$ (dashed black).\footnote{Since our description is non-relativistic, we ignore radiation. We also do not include baryons, hence the total matter density comes only from DM.} 

We can see that, initially, during the matter-dominated period, $\xi$ dominates the density budget of the universe, being responsible for almost all the matter density, while the contribution of $\chi$ remains small. As the matter density redshifts, the potential becomes increasingly important. Around the present time ($H_0 t \sim 1$), the interaction potential starts to dominate, resulting in a period of accelerated expansion. Note that the model does not solve the coincidence problem, because the onset of cosmic acceleration is set by the choice of parameters. 

The evolution can also be understood in the original $(\theta_1,\theta_2)$ variables. Cosmologically, the phonon fields are rotating along the symmetry-breaking direction with cosine modulations. In the early universe, the fields rotate rapidly, being oblivious to the small cosine modulations, and their kinetic energy dominates. Because of their non-canonical kinetic terms, this results in a DM-dominated universe. As the fields slow down due to Hubble expansion, they eventually feel the influence of the oscillatory potential, which triggers late-time acceleration. Meanwhile, on small, non-linear scales, the phonon gradient terms dominate again, resulting in DM behavior. This offers a compelling, unified picture for the dark sector of cosmology.

Coming back to Fig.~\ref{fig:densities_future}, for future times ($H_0 t \;\gsim\; 1$) we see deviations from the $\Lambda$CDM evolution, since our potential is dynamical and oscillatory. The density of $\xi$ decays from equality until the future due to the expected redshifting of a dust component, but also because of conversion from the ground state $\theta_1$ to the excited state $\theta_2$. Given this conversion, the density of $\chi$ starts to grow until it dominates the matter energy density, meaning that almost all the particles of the superfluid are in the excited state. As expected, the potential oscillates, becoming subdominant with respect to the DM component, around $7.5 \lesssim H_0 t \lesssim 9.5$, indicating a new phase of dust-like decelerated evolution, and then dominating the energy density again (for $H_0 t \;\gsim\; 9.5$) giving a new phase of accelerated expansion. Thus the future behavior of our model is quite distinct from that predicted by $\Lambda$CDM, showing an oscillation between epochs of decelerated, matter-dominated expansion and accelerated, potential-dominated expansion. We do not evolve this model past $H_0 t\simeq 10$, since after that time the null energy condition is violated and the effective field description breaks down.

\begin{figure}[htb]
\includegraphics[scale=0.48]{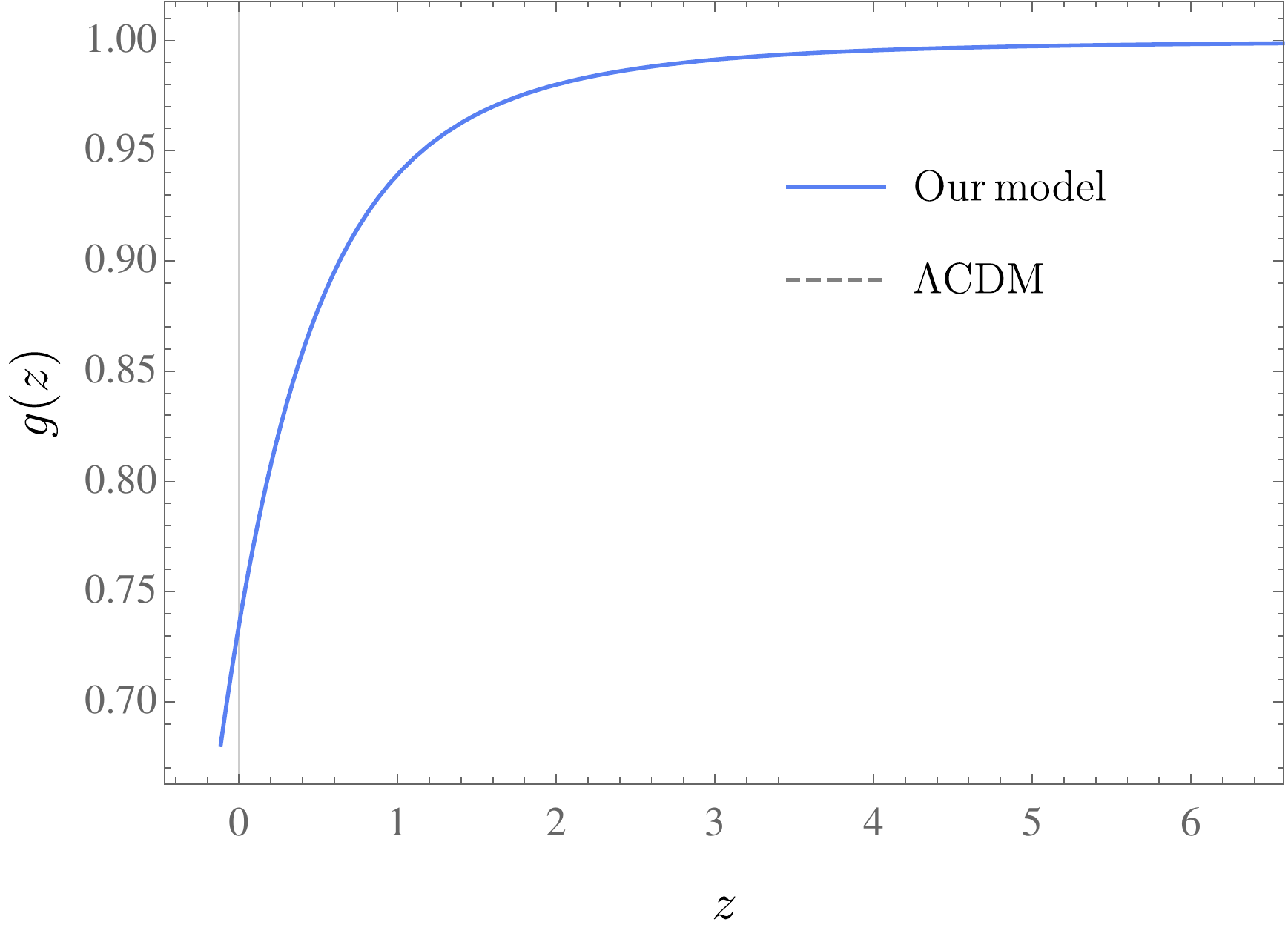}
\hfill
\includegraphics[scale=0.51]{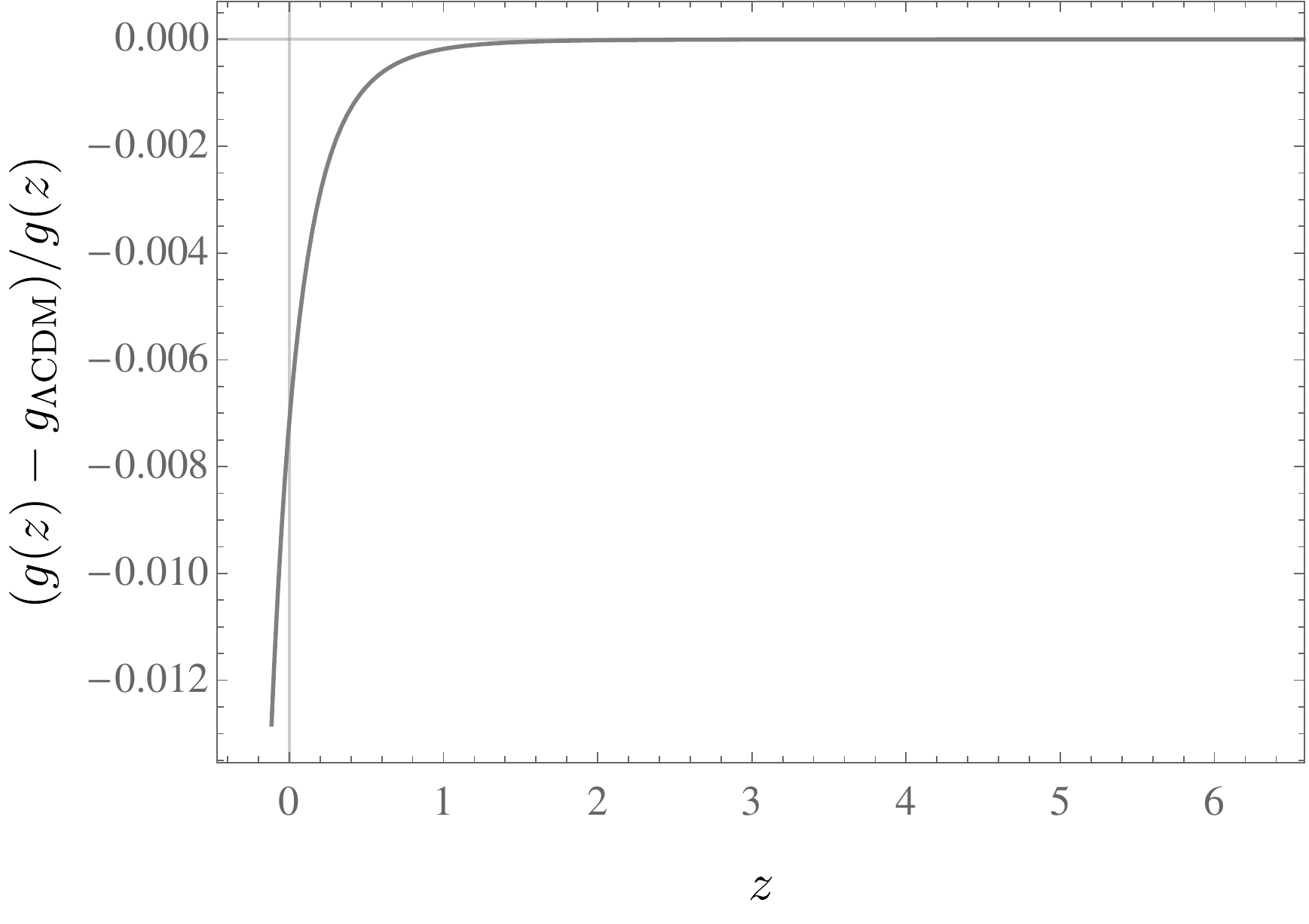}
\caption{\textit{Left:} Rescaled growth function $g(z) \sim (1+z) \delta(z)$ for our model (solid blue) and $\Lambda$CDM (dashed grey). We see that our predicted growth is somewhat suppressed at low redshift compared to $\Lambda$CDM. \textit{Right:} Fractional difference between our model and $\Lambda$CDM with the same matter density at early times and asymptotic Hubble constant $H_0$.
The difference is less than 2.5\% at $z = 0$ for our chosen parameters.}
\label{growth function}
\end{figure}

\subsection{Growth factor} 

We are finally in a position to solve~\eqref{2nd_order_eoms_final} for the evolution of linear density perturbations for our superfluid mixture.
From the form of these equations, notice that the interaction potential affects the growth in two ways: $i)$~through its impact on the background expansion,
{\it i.e.}, its contribution to $H(t)$; and $ii)$~its explicit appearance in the equation for~$\delta$. Deep in the matter-dominated epoch, however, the potential
is negligible, therefore $\delta$ evolve as in CDM. 

Let us say a word about our initial conditions, specified at matter-radiation equality. To match the observed primordial amplitude, we set $\delta_{\rm eq} = 10^{-5}$
at the initial time. Moreover, since $\delta$ evolves as a CDM perturbation in the matter-dominated era, we set its initial time derivative to the growing mode behavior $\delta \sim a$. In other words,
we set $\dot{\delta}_{\rm eq} = H_{\rm eq} \delta_{\rm eq}$.

With these initial conditions and the parameters mentioned at the outset, we numerically integrate~\eqref{2nd_order_eoms_final}. 
Figure~\ref{growth function} (left panel) shows the rescaled growth function, defined as\footnote{With this definition, standard CDM growth, $\delta \sim \frac{1}{1+z}$, corresponds to $g=1$.}
\beq
g(z) = \frac{1+z}{1+z_{\rm eq}} \frac{\delta(z)}{\delta_{\rm eq}}\,,
\eeq
for our model (solid blue curve) and $\Lambda$CDM model (dashed grey curve). Our model predicts a slightly suppressed growth compared to $\Lambda$CDM. As seen in the right panel of Fig.~\ref{growth function}, the difference is quite small, less than $1\%$ at $z=0$, for the parameters considered here. 

\begin{figure}[htb]
\includegraphics[scale=0.48]{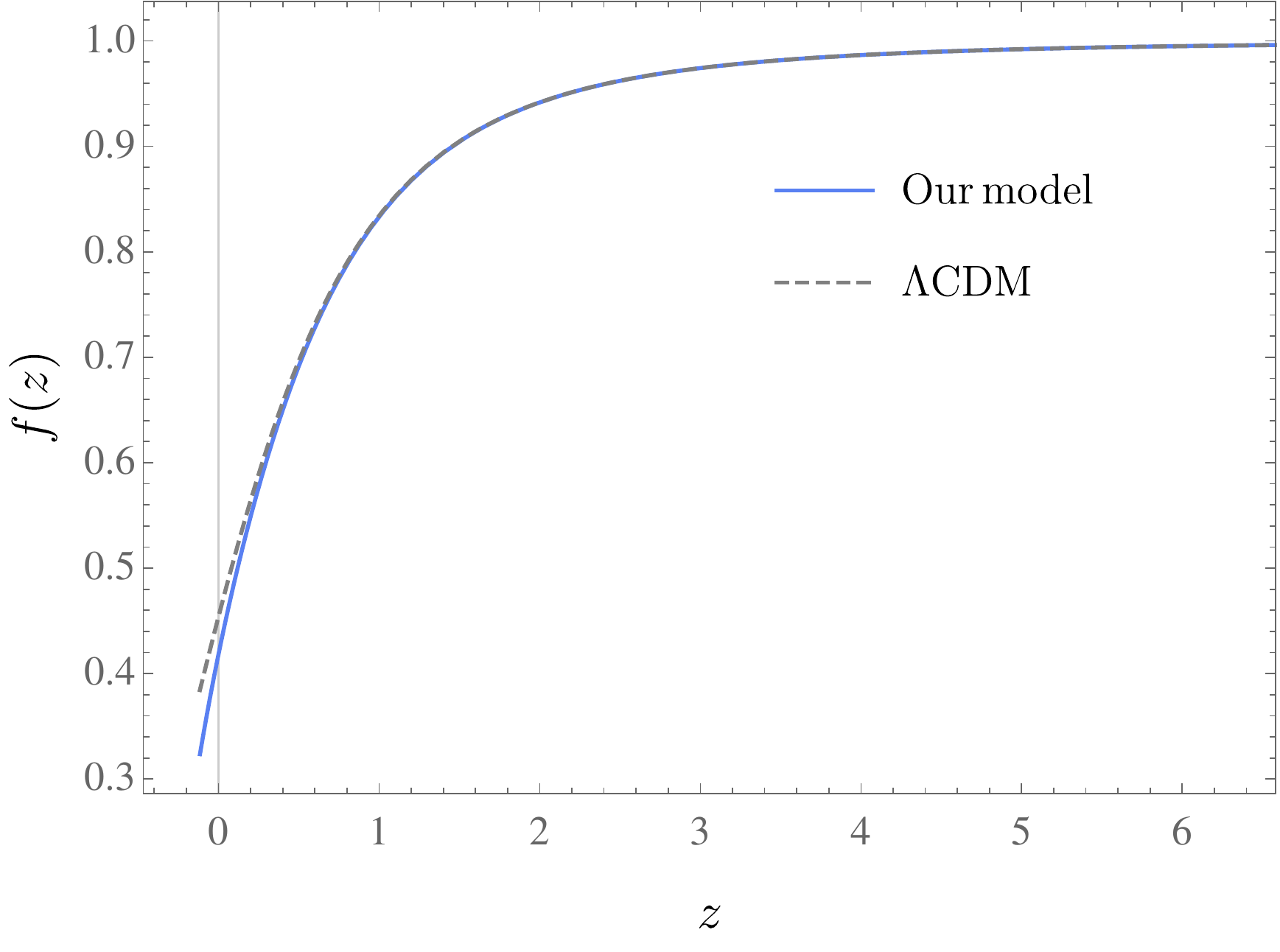}
\hfill
\includegraphics[scale=0.5]{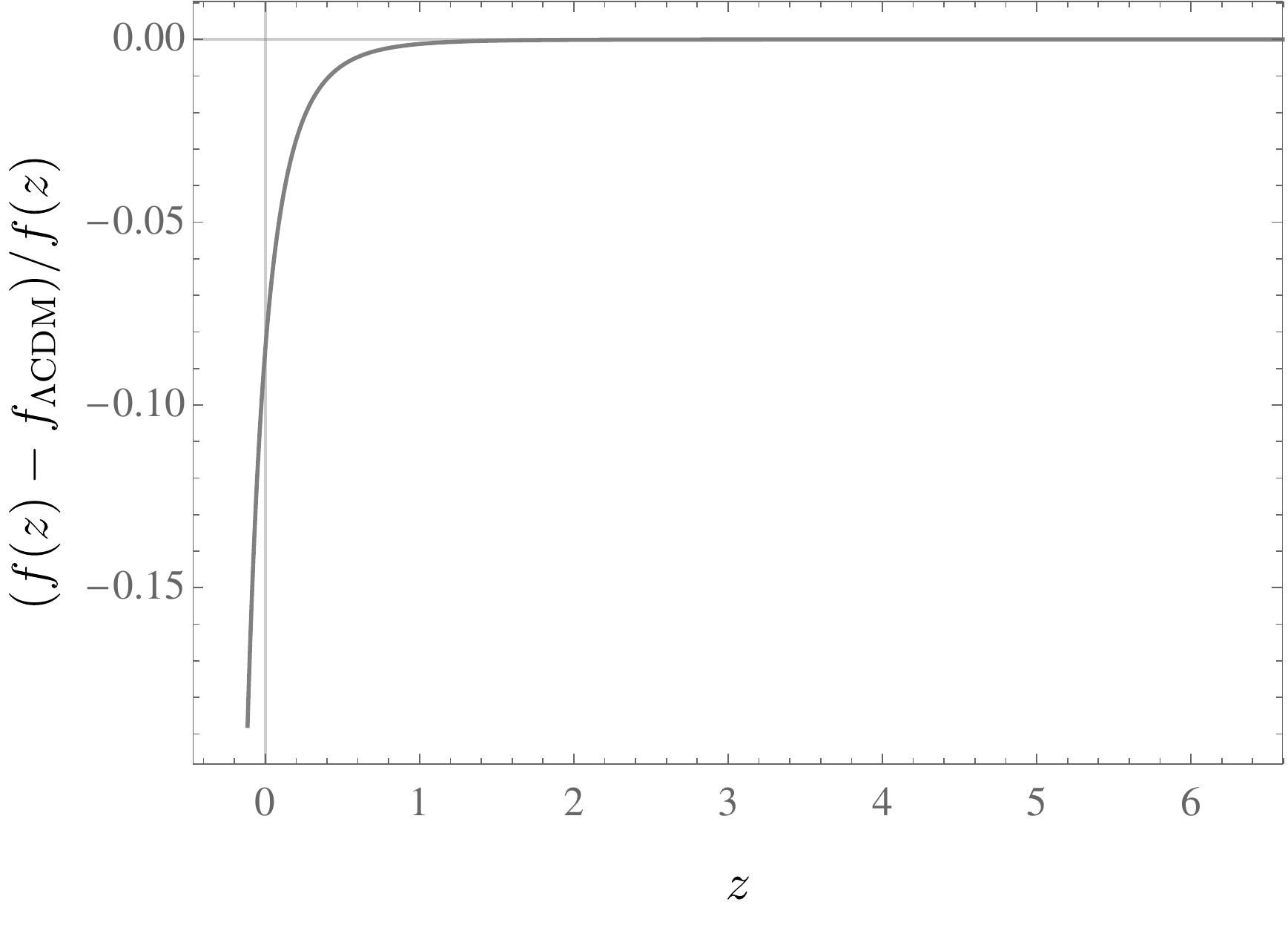}
\caption{\textit{Left:} Growth rate with respect to the redshift for our model (solid blue line), the prediction for $\Lambda$CDM (dashed gray line) and the expected growth rate using the Linder approximation with the expected growth index for $\Lambda$CDM (dotted red line). \textit{Right:} Fractional difference between our model and $\Lambda$CDM with the same matter density at early times and asymptotic Hubble constant $H_0$. The difference is substantial at $z = 0$ for our chosen parameters.}
\label{growth rate}
\end{figure}

The difference with $\Lambda$CDM is more pronounced with the growth rate, defined as 
\begin{equation}
    f(z) \equiv -\frac{{\rm d} \ln \delta(z)}{{\rm d} \ln (1+z)} = 1 - \frac{{\rm d} \ln g(z)}{{\rm d} \ln (1+z)}\,.
\end{equation}
This quantity is interesting since various probes of structure growth, such as redshift-space distortions, are sensitive to this quantity. 
In the left panel of Fig.~\ref{growth rate} we plot the evolution of the growth rate of our model (solid blue line), and, for comparison,
the expected growth rate for the $\Lambda$CDM model (gray dashed line) with the same asymptotic Hubble parameter $H_0$. Our
model deviates substantially, close to $10\%$, from the $\Lambda$CDM model, and predicts a steeper suppression of
density perturbation growth. We should caution that the quantitative difference depends of course on the parameter values chosen
here. In a future paper we plan to perform a more systematic study of model predictions and comparison to data, along the lines of~\cite{Costa:2013sva,Abdalla:2014cla}.

\section{Conclusion}
\label{conclusions}

In this paper we presented a novel unified framework for the dark sector, based on the idea of superfluid DM. The model relies on two non-relativistic superfluid species, which physically represent the ground state and excited state of DM. The two superfluid species interact through a Josephson/Rabi-like cosine potential, and this potential energy is responsible for driving late-time cosmic acceleration. The present theory nicely complements the recent proposal of DM superfluidity to explain the empirical success of MONDian phenomenology on galactic scales, thus offering a unified framework for DM, DE, and MOND. (A finite-temperature superfluid has also been shown recently to degravitate a large cosmological constant~\cite{Khoury:2018vdv}.)

Like pNGB models of DE, which also rely on a softly-breaking periodic potential, the model requires a super-Planckian decay constant to drive cosmic acceleration. Furthermore, thanks to the non-relativistic nature of the superfluids, their sound speeds remain small throughout the evolution and results in acceptable growth of density perturbations depending only on the expansion rate of the universe. This is unlike other earlier attempts of a unified dark sector, notably the Chaplygin gas models, where the adiabatic sound speed becomes relativistic at late times, resulting either in large oscillations or late-time growth in the matter power spectrum.

The interaction potential assumed in this work is motivated by the Josephson/Rabi form, which is ubiquitous in systems with multiple superfluid/superconducting species. Mixtures of BECs and superfluids with such interactions have already been created in the laboratory. This motivates us to study the implications of such potential for cosmology. Our study further emphasizes the 
potentially rich phenomenology of the superfluid framework, in particular the use of collective modes, in generating novel cosmological dynamics.

Leveraging the knowledge from condensed matter systems, many avenues offer themselves along which the present model could be further developed. 
An interesting future development for this theory could be to describe the Rabi interaction coefficient as a quantity that depends on temperature (or time) or even in the intra-particle distance, trying to give a dynamical explanation for the parameter $M$ that is liked to the onset of acceleration expansion. Since the theory was thus far described in the mean-field approximation at zero temperature, and the conversion is purely quantum, it would be interesting to study corrections to the mean-field approximation including finite-temperature effects. A first step in this direction was taken recently in~\cite{Sharma:2018ydn} in this case of a single component DM superfluid.

Our preliminary study of the predicted expansion history and linear growth of perturbations was done assuming a fiducial choice of parameter values, for illustrative purposes.
The small, yet significant deviations from $\Lambda$CDM (particularly for the growth rate) are tantalizing. They motivate us to explore more systematically the model predictions
over a broader range of parameters. In a future publication we plan to perform such a systematic analysis and compare the results to data. 

Although the numerical analysis focused primarily on the background expansion history and linear growth of perturbations, we derived analytically the general hydrodynamical equations governing the non-linear evolution of perturbations --- see~\eqref{fluid_eoms_inhomog}. We plan to solve these equations to study structure formation, for instance within the spherical collapse model. Along these lines, it would be interesting to study local effects coming from higher-temperature regions, such as galaxy clusters, where one might expect different relative populations for the species than the cosmological populations.

\appendix

\section{Derivation of the non-relativistic superfluid action from a Lorentz invariant theory} \label{Sec. Appendix}

 In this Appendix we show how the non-relativistic theory~\eqref{Ltheta1theta2} for two weakly-coupled superfluids can be derived from a Lorentz-invariant action. Our starting point is the relativistic theory of two complex scalar fields with quartic interactions:
\begin{equation}
\mathcal{L}_{0} = -\sum_{i = 1}^2 \sqrt{-g}\left( \left|\partial\Psi_i\right|^{2} + m_i^2 \left|\Psi_i\right|^{2}  + \frac{2m_i^4}{\Lambda^4_i} \left|\Psi_i\right|^{4}\right)\,.
\label{L0Phi's}
\end{equation}
where the dimensionless quartic couplings are chosen for later convenience. Decomposing the fields in polar variables,
\beq
\Psi_i = \frac{\rho_i}{\sqrt{2}} e^{{\rm i} \Theta_i} \,,
\eeq
the action can be written as
\begin{equation}
\mathcal{L}_{0} = -\sum_{i = 1}^2 \sqrt{-g}\left( \frac{1}{2}(\partial\rho_i)^2 + \frac{1}{2} \rho_i^2 (\partial\Theta_i)^2 +  \frac{1}{2}m_i^2 \rho_i^2+ \frac{m_i^4}{2\Lambda^4_i}\rho_i^4\right)\,.
\label{2scalarspolar}
\end{equation}

Thus far the two species are non-interacting, and the above theory enjoys a $U\left(1\right)\times U\left(1\right)$ global symmetry,
describing particle number conservation of each species separately. The conserved currents are
\beq
j^{\mu}_i=-\frac{{\rm i}}{\sqrt{2}}\left(\Psi^{*}_i\partial^{\mu}\Psi_i -\Psi_i\partial^{\mu}\Psi^{*}_i\right) = \rho_i^2 \partial^\mu \Theta_i \,; \qquad i = 1,2\,.
\label{ji}
\eeq
Below we will turn on a weak interacting potential that will break this symmetry down to a diagonal $U(1)$ subgroup. 

By definition, a superfluid is a phase of spontaneously broken $U(1)$ symmetry, at finite charge density $n_i  = j_{i\, 0} \sim \dot{\Theta}_i \neq 0$. 
Integrating out $\rho_i$ to leading order in derivatives, we obtain
\beq
\rho^2_i = \frac{\Lambda_i^4}{2m_i^4} \left(- \left(\partial\Theta_i\right)^2 - m_i^2\right)\,.
\label{rho2soln}
\eeq
For consistency, the state of interest must have $- \left(\partial\Theta_i\right)^2 \geq m_i^2$. The corresponding charge densities are
\beq
n_i = \frac{\Lambda_i^4}{2m_i^4} \left(- \left(\partial\Theta_i\right)^2 - m_i^2\right)\dot{\Theta}_i\,.
\label{nirel}
\eeq
Substituting~\eqref{rho2soln} back into~\eqref{2scalarspolar} gives the action for the phonon fields $\Theta_1$ and $\Theta_2$,
to leading order in the derivative expansion:
\beq
\mathcal{L}_{0} = \sum_{i = 1}^2 \sqrt{-g} \frac{\Lambda_i^4}{8}\left(\frac{\left(\partial\Theta_i\right)^2}{m^2_i}+1\right)^2\,.
\label{L0}
\eeq
The global $U(1)$ symmetries act non-linearly on the Goldstones as shift symmetries $\Theta_i \rightarrow \Theta_i + c_i$. 

At face value~\eqref{L0} is identical to the action for two ghost condensates~\cite{ArkaniHamed:2003uy}. However there are two important differences. The first difference pertains to the underlying description. Ghost condensation describes a fundamental scalar field, which spontaneously breaks Lorentz invariance. The superfluid effective description instead describes a collective degree of freedom, {\it i.e.}, sound waves. Secondly, the ghost condensate effective theory allows for field configurations with $\dot{\Theta}^2 < m^2$, resulting in violations of the Null Energy Condition. In the superfluid description, however, such a regime corresponds to negative particle number density, as can be seen from~\eqref{nirel}, and is therefore forbidden. 

We introduce now the Rabi-like coupling, given in the mean-field approximation by an interaction term that breaks the symmetry group to a residual global symmetry
$U\left(1\right)\times U\left(1\right)\rightarrow U\left(1\right)$:
\begin{equation}
\mathcal{L}_{\rm int}\propto -\frac{\Psi_1^{*}\Psi_2+\Psi_2^{*}\Psi_1}{\left|\Psi_1\right|\left|\Psi_2\right|}.
\label{interaction}
\end{equation}
In the non-relativistic regime, this becomes as oscillatory potential for the phonon difference:
\beq
\mathcal{L}_{\rm int} = - V(\Theta_2-\Theta_1) = -\frac{M^4}{2} \Big[ 1+\cos \left(\Theta_2-\Theta_1 \right) \Big]=-M^4\cos^2 \left(\frac{\Theta_2-\Theta_1}{2}\right)\,.
\label{LintTheta1Theta2}
\eeq
For convenience we have added a constant to the potential, which can come from adding a constant to the interaction term~(\ref{interaction}), such that the vacuum energy vanishes at the minimum of the potential.\footnote{Since we are not addressing the cosmological constant problem, we always have the freedom to shift the potential by an arbitrary constant.} The potential explicitly breaks the individual $U(1)$ symmetries down to the diagonal $U(1)$ subgroup that shifts the Goldstones by the same constant: $\Theta_i \rightarrow \Theta_i + c$. The charge densities $n_i$ are no longer separately conserved, but the total density, 
\beq
n = \sum_{i = 1}^2\frac{\Lambda_i^4}{2m_i^4} \left(- \left(\partial\Theta_i\right)^2 - m_i^2\right)\dot{\Theta}_i\,,
\label{ntotal}
\eeq
is conserved.

The Lagrangian is the sum of~\eqref{L0} and~\eqref{LintTheta1Theta2}. At this point we take the non-relativistic limit, writing the phonons as
\beq
\Theta_i = m_i t + \theta_i\,,
\eeq
with $\dot{\theta}_i \ll m_i$. Similarly the metric takes the weak-field, Newtonian form $g_{00} \simeq - (1+2\Phi)$, with $\Phi$ being the gravitational potential.

To leading order in $\dot{\theta}_i$, the densities $n_i$ for each species, given by~\eqref{nirel}, reduce to their non-relativistic expressions~\eqref{n_i non-rel}.
Furthermore, the Lagrangian reduces to
\beq
\mathcal{L} =  \frac{\Lambda_1^4}{2m_1^2} X_1^2 +\frac{\Lambda_2^4}{2m_2^2} X_2^2  - V(\theta_2-\theta_1 + \Delta E \,t)\,;\qquad X_i \equiv \dot{\theta}_i -m_i\Phi - \frac{(\vec{\nabla}\theta_i)^2}{2m_i}\,.
\label{Ltheta1theta2_appendix}
\eeq
This agrees with the non-relativistic action~\eqref{Ltheta1theta2} in the main text. Clearly, by generalizing the potential in~\eqref{L0Phi's} for $\Psi_1$ and $\Psi_2$, we can obtain different $P(X_i)$ superfluid kinetic terms, and arrive at~\eqref{eq:model} through similar steps. The procedure for obtaining a general $P(X)$ theory, as well as particular cases like k-essence, DBI and ghost condensate, can be found in the literature, \textit{e.g.} \cite{Babichev:2018twg,Babichev:2017lrx,Tolley:2009fg,Bilic:2008pe,Bilic:2008zk}. An important difference from what is done in the literature to our case is that our theory is not described in terms of fundamental scalar fields, but using phonons or collective superfluid modes, so fluid caustics in our model are naturally resolved, in contrast with the situation in the aforementioned models.

\begin{acknowledgments}
We are grateful to Tami Pereg-Barnea for many helpful conversations on coupled superfluids. We thank Lasha Berezhiani, Anushrut Sharma and Eduardo Matsushita for helpful conversations. The work at McGill University is supported in part by an NSERC Discovery Grant and by funds from the Canada Research Chair program.  The work of J.K. is supported in part by the US Department of Energy (HEP) Award DE-SC0013528, NASA ATP grant 80NSSC18K0694, the Charles E. Kaufman Foundation of the Pittsburgh Foundation, and a W.~M.~Keck Foundation Science and Engineering Grant.  E.\,F. and G.\,F. also acknowledge financial support from CNPq (Science Without Borders).
\end{acknowledgments}


\end{document}